\documentclass[twocolumn,times,tighten]{aastex63}

\received{\textit{2019 October 21}}
\revised{\textit{2019 November 27}}
\accepted{\textit{2019 November 27}}
\submitjournal{\textit{(and accepted by) The Astronomical Journal}}
\shorttitle{Uranus in Mid-Spring: Temperatures and Circulations}
\shortauthors{Roman et al.}

\begin{document}

\title{Uranus in Northern Mid-Spring: Persistent Atmospheric Temperatures and Circulations Inferred from Thermal Imaging}

\correspondingauthor{Michael T. Roman}
\email{mr359@le.ac.uk}

\author[0000-0001-8206-2165]{Michael T. Roman}
\affil{University of Leicester \\
School of Physics and Astronomy \\
University Road, Leicester, LE1 7RH, UK}

\author[0000-0001-5834-9588]{Leigh N. Fletcher}
\affil{University of Leicester \\
School of Physics and Astronomy \\
University Road, Leicester, LE1 7RH, UK}

\author[0000-0001-7871-2823]{Glenn S. Orton}
\affil{Jet Propulsion Laboratory, MS 183-501, Pasadena, CA 91109, USA \\}

\author[0000-0001-8692-5538]{Naomi Rowe-Gurney}
\affil{University of Leicester \\
School of Physics and Astronomy \\
University Road, Leicester, LE1 7RH, UK}

\author[0000-0002-6772-384X]{Patrick G. J. Irwin}
\affil{University of Oxford \\
Atmospheric, Oceanic, and Planetary Physics, Department of Physics\\ Clarendon Laboratory, Parks Road, Oxford OX1 3PU, United Kingdom}



\begin{abstract}

We present results from mid-infrared imaging of Uranus at wavelengths of 13.0 $\mu$m and 18.7 $\mu$m, sensing emission from the stratosphere and upper troposphere, acquired using the VISIR instrument at the Very Large Telescope (VLT), September 4\textendash October 20, 2018.  Using a combination of inverse and forward modeling, we analyze these northern mid-spring (L$_s$$\sim$46$^{\circ}$) images and compare them to archival data to assess seasonal changes since the 1986 southern solstice and subsequent equinox. We find the data are consistent with little change ($<$ 0.3 K) in the upper tropospheric temperature structure, extending the previous conclusions of \cite{orton2015thermal} well past equinox, with only a subtle increase in temperature at the emerging north pole. Additionally, spatial-temporal variations in 13-$\mu$m stratospheric emission are investigated for the first time, revealing meridional variation and a hemispheric asymmetry not predicted by models. Finally, we investigate the nature of the stratospheric emission and demonstrate that the observed distribution appears related and potentially coupled to the underlying tropospheric emission six scale heights below. The observations are consistent with either mid-latitude heating or an enhanced abundance of acetylene. Considering potential mechanisms and additional observations, we favor a model of acetylene enrichment at mid-latitudes resulting from an extension of the upper-tropospheric circulation, which appears capable of transporting methane from the troposphere, through the cold trap, and into the stratosphere for subsequent photolysis to acetylene. 

\end{abstract}


\section{Introduction} \label{sec:intro}

The atmosphere of Uranus is subject to a unique pattern of seasonal forcing.  Due to the planet's ~98$^{\circ}$ obliquity, nearly all latitudes on Uranus experience seasonally extended periods of total daylight and darkness.  Averaged over a full orbital period, this seasonal cycle results in more solar energy being deposited annually at the poles than at the equator, contrary to the other solar system planets \citep{friedson1987seasonal,conrath1990temperature,moses2018seasonal}. While observations of reflected light clearly show a seasonal cycle in the tropospheric hazes at polar latitudes \citep{[e.g.] karkoschka2001uranus, rages2004evidence, irwin2010revised, roman2018aerosols, toledo2018uranus,sromovsky2019methane,lockwood2019final}, the effects of this peculiar seasonal forcing  on Uranus' temperature field, atmospheric chemistry, and circulations have yet to be fully observed or modeled. 

Given the pattern of solar forcing, early radiative-convective modeling suggested the summer pole of Uranus should be $\sim$1.5\textendash6 K warmer than the equator and winter pole \citep{wallace1983seasonal,friedson1987seasonal}. The first and most complete measurements of the pole-to-pole variation in thermal emission from Uranus came in 1986 with the Infrared Interferometer Spectrometer and Radiometer (IRIS) \citep{hanel1986infrared} aboard Voyager 2. Spectral data from the flyby were inverted to yield upper tropospheric (70\textendash 400 mbar) temperatures near the time of the southern summer solstice.  Contrary to expectations, these data showed the warmest temperatures at the equator and poles, with colder temperatures at mid-latitudes \citep{flasar1987voyager,conrath1990temperature,orton2015thermal}, interpreted by \citet{flasar1987voyager} as an indication of an organized atmospheric circulation, with upwelling at mid-latitudes and downwelling at the equator and high-latitudes producing adiabatic cooling and heating, respectively.  Although the coldest portions of the northern winter hemisphere were slightly colder ($\sim$1 K) than the equivalent latitudes in the summer hemisphere, the two poles were roughly the same temperature at the tropopause, and the summer pole was inferred to be only marginally warmer in the lower stratosphere \citep{orton2015thermal}. 

\cite{flasar1987voyager} and \cite{conrath1990temperature} proposed that the inferred lack of seasonal response could be largely explained by atmospheric radiative time constants ($t_r$) that were long relative to Uranus orbital period ($t_o$), specifically such that 2$\pi$ $t_r$$/$$t_o$ $>>$1. Using a radiative-convective-dynamical model, \cite{conrath1990temperature} suggested that a radiative time constant of 130 years would appropriately dampen the atmospheric temperature response by a factor of 10, resulting in seasonal variation of only $\sim$1 K, lagging roughly a full season behind the instantaneous solar forcing. They noted, however, that this would not explain the hemispheric asymmetries, inconsistent with symmetric equinoctial forcing, present in the seasonally-lagged solstitial data. 

One season later, around the time of 2007 southern autumnal equinox (L$_s$$\sim$0$^{\circ}$), Uranus' spatially resolved radiance was evaluated using thermal ground-based imaging and compared to the earlier Voyager measurements by \cite{orton2015thermal}.  The study found very little if any change ($<$ 0.4 K) in the implied thermal structure from southern summer solstice to southern autumnal equinox, including the noted asymmetry at mid-latitudes, from which it was concluded that upper tropospheric radiative time constants must be no less than $\sim$ 330 years. At that value, the phase lag would have essentially reached a maximum of roughly a season as the amplitude of seasonal temperatures variations diminishes \citep{conrath1990temperature}. 

The long radiative time constants suggested by the data are apparently in conflict with values calculated using radiative-transfer modeling. Theoretical radiative time constants as a function of height for the outer planets atmospheres have been re-evaluated recently by \cite{li2018high} using a radiative transfer model with improved correlated-k gas opacities and state-of-the-art photochemical models. For Uranus, \cite{li2018high} calculated radiative time constants with values less than \cite{conrath1990temperature}, ranging from roughly 15 to 70 years at 400 to 70 mbar pressures.  The reason for this apparent discrepancy between theoretically predicted values and those inferred from observations are unknown,  but it may possibly indicate errors in the assumed abundances, distribution, and/or opacities of radiatively active hydrocarbons. 

The distribution of hydrocarbons produced photochemically from the seasonally varying solar flux on Uranus has been recently modeled by \cite{moses2018seasonal}. Though chemical abundances were constrained by disk-averaged spectra of Uranus from \textit{Spitzer} at northern vernal equinox \citep{orton2014mid}, the theoretically predicted seasonal variations remain unverified by data.  Likewise, the spatial distributions of hydrocarbon species remain largely unknown given the nearly complete absence of constraining spatially resolved images and spectra in the literature. As \cite{moses2018seasonal} notes, their model does not include the potentially significant role of the general circulation, which could potentially produce gradients in the methane abundance and consequent photolytic products. As we will show, meridional variation in hydrocarbons or temperature are indeed required for modeling the observed disk-resolved emission associated with acetylene, suggesting a significant dynamical link between the troposphere and stratosphere extending over several pressure scale heights. 

To this end, we present and analyze newly acquired mid-infrared images of Uranus\textemdash halfway through Uranus' northern spring with the north pole now fully in view for the first time in the era of thermal imaging \textemdash to investigate temporal changes in atmospheric temperatures and circulation.  New and archival data are introduced in Section \ref{sec:data}, followed by a description of our methods for analysis in Section \ref{sec:methods} and results in Section \ref{sec:results}. We conclude with a discussion of the implications of our results in Section \ref{sec:discussion} and brief summary of conclusions in Section \ref{sec:conclusions}. 

\section{DATA} \label{sec:data}
\subsection{New Observations: 2018 images}
Images of Uranus were acquired in September and October of 2018 using the mid-infrared VISIR instrument \citep{lagage2004visir,kerber2014visir} at the European Southern Observatory's (ESO) Very Large Telescope (VLT, UT3).  Two separate filters were used: Q2, with a central wavelength of 18.7 $\mu$m (534.2 cm$^{-1}$) and a full width at half maximum width of 0.88 $\mu$m; and NeII\_2, centered at 13.0 $\mu$m (769.2 cm$^{-1}$) with a width of 0.22 $\mu$m.

\begin{figure*}[ht!]
\includegraphics[width=\linewidth, trim=.0in 0in 0in .0in, clip]{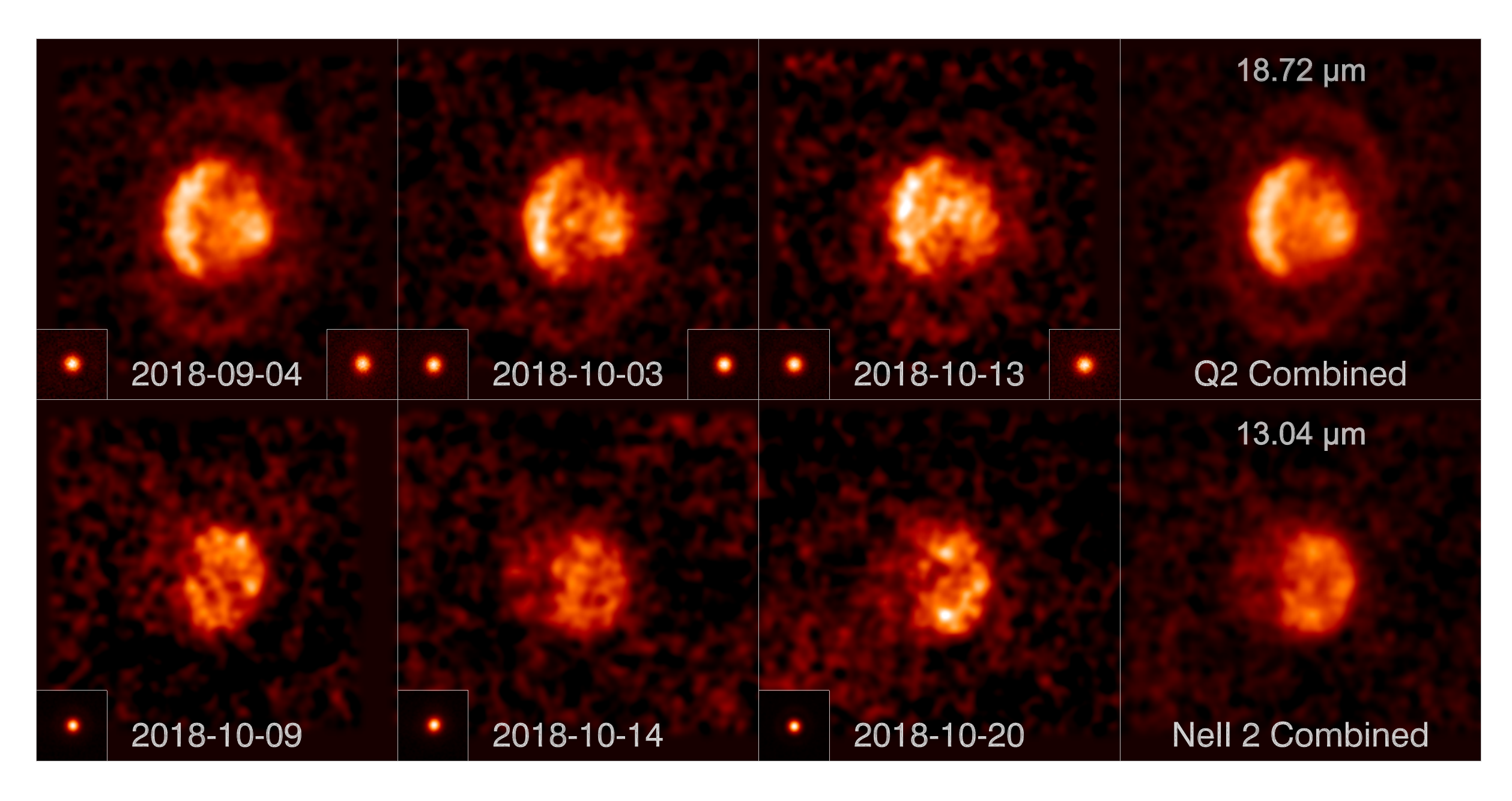}
\caption{2018 VLT-VISIR images of Uranus in two different filters: Q2 (top row), with a central wavelength of 18.72 $\mu$m, and NeII\_2 (bottom row), centered at 13.04 $\mu$m. These images were acquired over separate nights (dates indicated), and, for each filter, results from individual nights were averaged together, weighted by their noise, to produce the combined images. For clarity, the images are shown following a low-pass filtering in the form of a Gaussian blur with a standard deviation of three pixels. Insets show the stars used for calibration and are representative of the spatial resolution of the seeing disc. Details for these images are provided in Table \ref{table:obstable}.  Note the emission from Uranus' rings (primarily the $\epsilon$ ring) measured at 18.72 $\mu$m for the first time; for a thermal analysis of rings including these data, see \cite{molter2019thermal}. \label{fig:obs}}
\end{figure*}

The 18.7-$\mu$m filter (Q2) is sensitive to the continuum emission from atmospheric hydrogen and thus serves as a probe of upper tropospheric temperatures with a peak contribution from 200 mbars at nadir geometry (contribution functions are provided in Appendix Figure \ref{fig:filters}). Images show enhanced emission at the equator and pole relative to mid-latitudes, qualitatively consistent with inferences of mid-latitudinal cooling in the upper troposphere dating back to the Voyager era \citep{flasar1987voyager,conrath1990temperature,orton2015thermal}. 

In contrast, the 13.0-$\mu$m filter (NeII\_2) is dominated by thermal emission associated with stratospheric acetylene with maximum contributions from pressures near 0.2 mbar. The 13.0-$\mu$m images show brighter emission at middle and high latitudes compared to the equator.  Since C$_2$H$_2$ is a minor and likely variable species \citep{moses2018seasonal}, variation in the 13.0-$\mu$m emission can be attributed to variation in either the C$_2$H$_2$ abundance, the stratospheric temperatures, or a combination of the two quantities.

Jointly, these images reveal tropospheric temperatures and stratospheric emission from Uranus' northern hemisphere in mid-spring (sub-solar latitudes of 44-45$^{\circ}$), including unprecedented views of Uranus' north pole and the first-ever images of Uranus' rings in the mid-infrared. Our analysis of the rings based on these images is presented in \cite{molter2019thermal}. 
Data were collected in $\sim$45 minute blocks on six different nights (three nights for each filter) and reduced using ESO's pipeline of standard infrared chopping and nodding techniques.  The resulting images (see Figure \ref{fig:obs}) were flux calibrated via comparison to observed standard stars. We assumed a systematic error of up to 30\% in radiance due to fluctuations in sky brightness and up to 20\% due to uncertainties in the stellar fluxes \citep{dobrzycka2008calibrating}. Random errors were estimated from the standard deviation of the background sky. To improve the signal-to-noise ratio (SNR) in each filter, the calibrated images from each night were weighted by the inverse of their squared random errors and combined to yield the final, absolutely calibrated mean image for each filter. A summary of our 2018 observations is provided in Table \ref{table:obstable}.

The random noise was significant in both images given the weak signal from Uranus' cold atmosphere. For individual images, we estimate an average signal-to-noise ratio (SNR) for individual pixels upon the disk to be as low as $\sim$1.2 at 18.7 $\mu$m and $\sim$0.9 at 13.0 $\mu$m.  Combining the nights increased these SNRs to $\sim$2.8 and $\sim$1.6, respectively, varying with signal across the disk.  Zonal averaging and meridional binning were employed to significantly enhance the SNR in computed latitudinal profiles.

\begin{deluxetable*}{cccccc}
\centering
\tablecolumns{6}
\tablecaption{Summary of 2018 Observations}
\tablehead{
\colhead{Filter [Wavelength]} & \colhead{Date (UT)} & \colhead{Time (UTC)} & \colhead{Airmass} & \colhead{Calibration Star} & \colhead{Sub-Observer (Solar) Latitude}}
\startdata
Q2 [18.72 $\mu$m] & 2018-09-04 & 8:14 - 9:03 & 1.249 - 1.316 & HD013596, HD009692 & 45.3$^{\circ}$ (43.0$^{\circ}$)\\
& 2018-10-03 & 5:15 - 5:54 & 1.236 - 1.255 & HD008498, HD010380 & 44.4$^{\circ}$ (43.4$^{\circ}$)\\
& 2018-10-13 & 5:32 - 6:04 & 1.240 - 1.272 & HD011353, HD040808 & 44.0$^{\circ}$ (43.5$^{\circ}$)\\
\\
NeII\_2 [13.04 $\mu$m] & 2018-10-09 & 5:13 - 5:55 & 1.238 - 1.245 & HD008498 & 44.2$^{\circ}$ (43.4$^{\circ}$)\\
 & 2018-10-14 & 4:51 - 5:39 & 1.234 - 1.237 & HD008498 & 44.0$^{\circ}$  (43.5$^{\circ}$)\\
 & 2018-10-20 & 4:49 - 5:36 & 1.233 - 1.271 & HD011353 & 43.8$^{\circ}$ (43.5$^{\circ}$)\\
\enddata
\tablecomments{The VISIR instrument in 2018 had a plate scale of 0.0453 arcsecs/pixel.}
\label{table:obstable}
\end{deluxetable*}

\begin{deluxetable}{ccccc}
\centering
\tablecolumns{4}
\tablecaption{Summary of 2009 Observations}
\tablehead{
\colhead{Filter [Wavelength]} & \colhead{Date (UT)} & \colhead{Time (UTC)} & \colhead{Airmass}}
\startdata
Q3 [19.50 $\mu$m] & 2009-08-03 & 05:40 - 06:12 & 1.164 - 1.247\\
&  & 06:20 - 06:53 & 1.104 - 1.151\\
&  & 06:56 - 07:19 & 1.086 - 1.102\\
&  & 07:45 - 08:10 & 1.082 - 1.091\\
&  & 08:21 - 08:44 & 1.098 - 1.125\\
&  & 08:46 - 09:10 & 1.167 - 1.152\\
&  & 09:13 - 09:35 & 1.173 - 1.272\\
&  & 09:38 - 10:01 & 1.237 - 1.229\\
\\
NeII\_2 [13.04 $\mu$m] & 2009-08-05 & 06:56 - 07:19 & 1.083 - 1.095\\
 &  & 07:32 - 07:56 & 1.081 - 1.087\\
 &  & 08:10 - 08:33 & 1.097 - 1.122\\
 &  & 08:46 - 09:10 & 1.142 - 1.188\\
 &  2009-08-06 & 07:26 - 07:59 & 1.081 - 1.091\\
 &  & 08:02 - 08:32 & 1.092 - 1.125\\
 &  & 08:39 - 09:13 & 1.132 - 1.201\\
 &  & 09:15 - 09:49 & 1.207 - 1.323\\
\enddata
\tablecomments{The VISIR instrument in 2009 had a plate scale of 0.0750 arcsecs/pixel. Standard stars were observed for PSF determination, but photometric calibration was performed via comparison Spitzer values \citep{orton2014a}. HD12524, HD5112, and HD216149 were observed in the NEII\_2 filter; HD224630, HD220954, HD220009, and HD198048 were observed in the Q3 filter. For all images, the sub-observer latitude was $\sim$8.8$^{\circ}$, and the sub-solar latitude was $\sim$6.7$^{\circ}$ }
\label{table:equitable}
\end{deluxetable}

\begin{figure}[ht!] 
\includegraphics[width=\linewidth, trim=.4in 1.6in .4in 1.6in, clip]{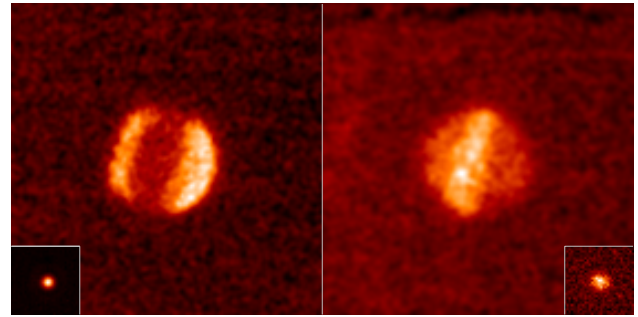}
\caption{ Mean near-equinoctial VLT-VISIR observations of Uranus \citep{orton2018neii} used for our retrievals to evaluate temporal changes in the 13.0-$\mu$m data. (Left) Mean NeII\_2 (13.04-$\mu$m) image from August 5th and 6th, 2009. (Right) Mean Q3 (19.5-$\mu$m) image from August 3, 2009, sensing upper-tropospheric depths similar to the Q2 (18.7-$\mu$m) images.}
\label{fig:equiobs}
\end{figure}

\subsection{Archival Data} \label{sec:archival}
To evaluate temporal changes in the 18.7-$\mu$m images, we compared our new data to Voyager data (the oldest spatially resolved thermal data of Uranus available) following the techniques of \cite{orton2015thermal}, as described in Section \ref{sec:methods}.  In the case of the 13.0 $\mu$m stratospheric emission, the Voyager IRIS instrument was unfortunately not sensitive enough to provide measurements, and so changes in the NeII\_2-band images were evaluated using VLT-VISIR images dating from near equinox \citep{orton2018neii} (see Figure \ref{fig:equiobs}). 

\begin{figure}[ht!] 
\includegraphics[width=\linewidth, trim=0in 0in 0in 0in, clip]{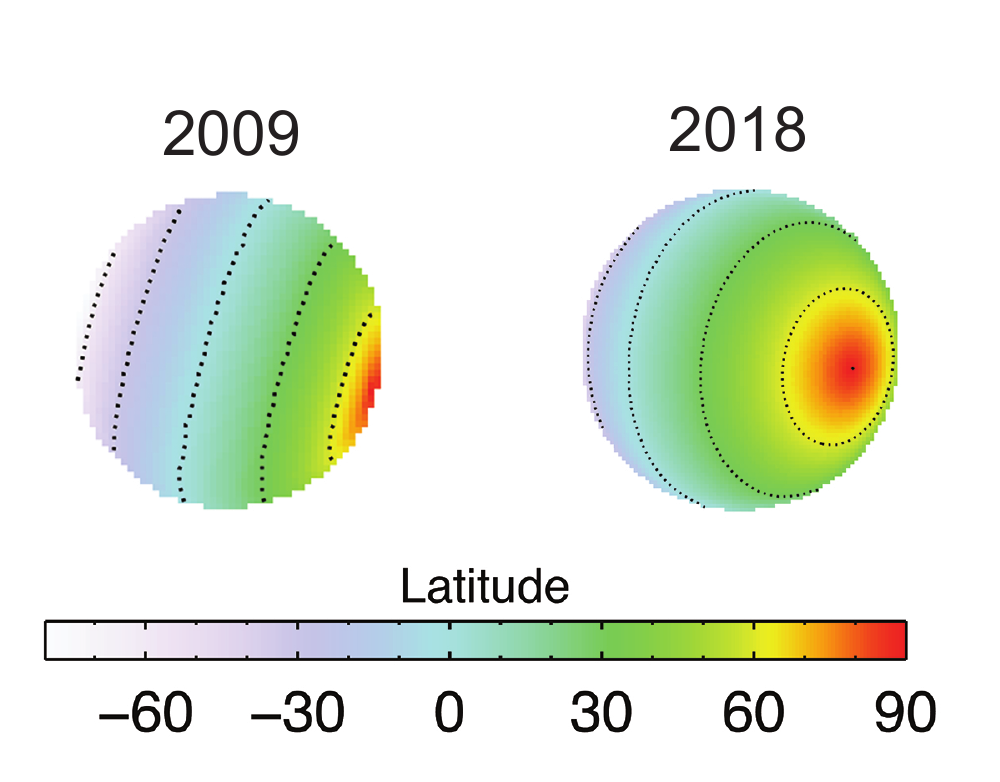}
\caption{Latitudinal coverage for the 2009 (left) and 2018 (right) observations. Latitudes are indicated by the color bar.  Dotted lines additionally mark latitudes of -60$^{\circ}$, -30$^{\circ}$, 0$^{\circ}$, 60$^{\circ}$, and 90$^{\circ}$. The 2018 map has a finer pixel resolution relative to the 2009 map due to the improved plate scale.}
\label{fig:latimages}
\end{figure}

The near-equinoctial data included two consecutive nights of 13-$\mu$m imaging from 2009, but no contemporaneous 18.7-$\mu$m images were available. We instead substituted a similarly sensitive 19.5-$\mu$m image (Q3 filter) to define the tropospheric emission when performing the stratospheric retrievals discussed in Section \ref{sec:methods}. Due to a previously greater plate scale on VISIR  (0.075 versus 0.0453 arcsecs per pixel), 2009 images were acquired at a courser spatial resolution than the 2018 images, but they otherwise present equivalent measurements at a different time and viewing geometry. The latitudinal coverage and relative plate scale for both the 2009 and 2018 observations can be seen in Figure \ref{fig:latimages}. 

Data from each night were collected in blocks, reduced using the ESO pipeline, and combined to yield nightly averages.  We initially attempted to calibrate these images using standard calibration stars, but we found resulting radiances to be 10-40\% less than the disk-integrated values from the more reliable Spitzer data acquired less than two years earlier \citep{orton2014a}. Ultimately, given uncertainty in the ground-based calibrations and the relative contemporaneity of the space-based Spitzer observations, we chose to scale the 2009 images such that their disk integrated radiances equaled the Spitzer values at the equivalent central wavelengths. Details on these observations are found in Table \ref{table:equitable}.

The 2009 images reveal tropospheric temperatures and stratospheric emission near equinox (L$_s$ $\sim$7), with sub-solar latitudes of $\sim$6.7$^{\circ}$. Signal-to-noise ratios were $\sim$6-7 for the 19.5-$\mu$m and $\sim$3-4 for the 13.0-$\mu$m images. Though the systematic uncertainties for Spitzer are less than 10\%, we conservatively estimate systematic errors as high as 20\% to account for changes in viewing geometry and radiance in the intervening 1.5 years.

\section{Methods} \label{sec:methods}
\subsection{Comparing Voyager Tropospheric Temperatures to Q2 Images}
To evaluate temporal changes in the tropospheric emission, we followed the method of \cite{orton2015thermal}, which compared synthetic images, derived from Voyager-IRIS spectra, to a collection of ground-based imaging data.  In that work, Q-band images (including 18.7 $\mu$m and 19.5 $\mu$m VISIR data from 2006 and 2009, respectively,) were analyzed to assess changes in the meridional distribution of upper tropospheric temperatures between the 1986 Voyager encounter (roughly coinciding with Uranus' southern summer solstice) and the following equinox (southern autumn) in 2007. No significant changes were detected \citep{orton2015thermal}. Using a similar approach, we compared our 2018 images to the 1986 Voyager data to evaluate potential changes over a longer interval.

First, we forward-modelled emission from the temperatures inferred from the Voyager measurements to produce a synthetic image of Voyager-era emission with 2018 geometry\textemdash effectively what we would have observed at 18.7 $\mu$m in 2018 if the upper tropospheric temperatures near 200 mbar remained unchanged since 1986. These temperatures were taken directly from \cite{orton2015thermal}, which reanalyzed Voyager 2-IRIS spectra acquired over the four days surrounding the closest approach to Uranus on January 24, 1986. These data covered both the southern and northern hemispheres as the spacecraft passed Uranus at distances ranging from 4.2 to 112 Uranus radii. Spectra were inverted to yield temperatures as function of latitude and pressure from 70-400 mbar and then extended to lower pressures based on Spitzer observations \citep{orton2014a}. We refer the reader to \cite{orton2015thermal} for further details on how the Voyager temperatures were retrieved. 

We then convolved the forward-modelled images of radiance with point spread functions (PSF) determined from the corresponding stellar images to model the effects of atmospheric seeing and instrumental diffraction.  To account for any errors associated with imperfect navigation and stacking of our three images, we created a synthetic image for each of the three individual observations and combined them in precisely the same way as we had done for the data. Finally, zonal averages (binned 10$^{\circ}$ in latitude to improve the SNR of the data) were extracted from both the synthetic and real images, converted from units of radiance into brightness temperatures, and compared as a function of latitude (\textit{e.g.}, see Figures \ref{fig:datvmod}, \ref{fig:2018vsVoy}).

Since \cite{orton2015thermal} found no significant changes in the upper tropospheric temperatures between the 1986 solstice and the 2007 equinox (using a wider range of VISIR data and the same radiative transfer code applied in the present study), we chose to only assess changes between the 1986 Voyager data and our 2018 18.7-$\mu$m data.

\subsection{Evaluating Stratospheric Changes}
Evaluating temporal changes at the roughly 0.2-mbar heights sensed by the 13.0-$\mu$m (NeII\_2) filter (see contribution functions in Appendix Figure \ref{fig:filters}) required a more complicated approach combining radiative-transfer forward modeling and retrievals. Since the Voyager IRIS measurements were insensitive to the weak emission at 13 $\mu$m, we drew upon archival 13-$\mu$m images from 2009 for comparison (see Section \ref{sec:archival}).  However, given the nine years difference between the 2018 and 2009 images, a direct, quantitative comparison required accounting for differences in the observing geometry that can affect the observed radiance as a function of emission angle. To account for this, we took the following approach: First, starting with a prior 1-D model of \cite{orton2014a}, we retrieved zonally averaged temperature and acetylene profiles from the observations, accounting for the dependence on emission angle, and used these retrieved values to create a model of the atmosphere (\textit{i.e.,} the temperature and acetylene as a function of pressure and latitude). Then, as we had done with the Voyager temperatures, we computed the radiances emitted from this atmospheric model at different desired emission angles, resulting in a synthetic image (i.e., a forward model) of the planetary emission at a different viewing geometry. 

In practice, gradients in the observed emission in our data can be equally explained by gradients in the stratospheric temperatures or the acetylene mixing ratios. We performed retrievals in which one of the two parameters (either temperature or acetylene) was held fixed above the 1-mbar height (assuming values of \cite{orton2015thermal} and \cite{moses2018seasonal}) while the other parameter was free to vary.  We found 1 mbar to be a reasonable pressure boundary for separating the 13-$\mu$m filter contribution from the deeper 18.8-$\mu$m contribution. Although the 13-$\mu$m contribution function showed some sensitivity to pressures greater than 1 mbar, enhancement in parameters at these deeper levels were found to produce excessive limb darkening, strongly inconsistent with the center-to-limb variation in the observations.  Regardless, retrieved models of either stratospheric temperatures or acetylene worked equally well for forward modelling the emission.

\begin{figure*}[ht!]
\includegraphics[width=\linewidth, trim=.6in 3.2in 1in 1.8in, clip]{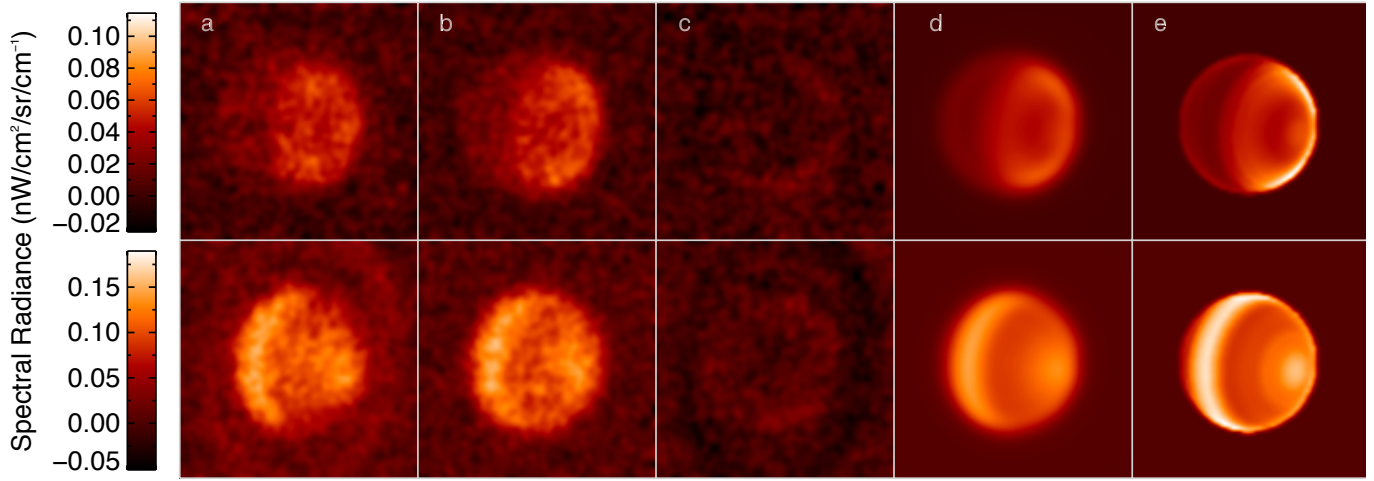}
\caption{Comparison of 2018 data and synthetic images at 13.0 $\mu$m (NeII\_2, top) and 18.7 $\mu$m (Q2, bottom).  Synthetic images were produced from our model based on temperatures retrieved from the data.  From left to right, the figure shows a) real data; b) our synthetic images convolved with a PSF to simulate diffraction and atmospheric distortion, with synthetic noise added; c) the model minus the data showing only slightly excessive limb brightening in our models; d) the convolved synthetic image without noise; e) the forward modeled synthetic disk prior to any degradation. The latter represents an idealized model of unadulterated emission from Uranus that is consistent with observations; the finest banding structure and contrast exceed the spatial resolution of the data and should not be regarded as physically significant.  Zonal averages extracted from panels $d$ and $e$ correspond to the solid and dashed lines in Figure \ref{fig:datvmod}, respectively. \label{fig:imgmod2018}}
\end{figure*}

\begin{figure*}[ht!]

\includegraphics[width=\linewidth, trim=.7in 2.4in 1in 2.5in, clip]{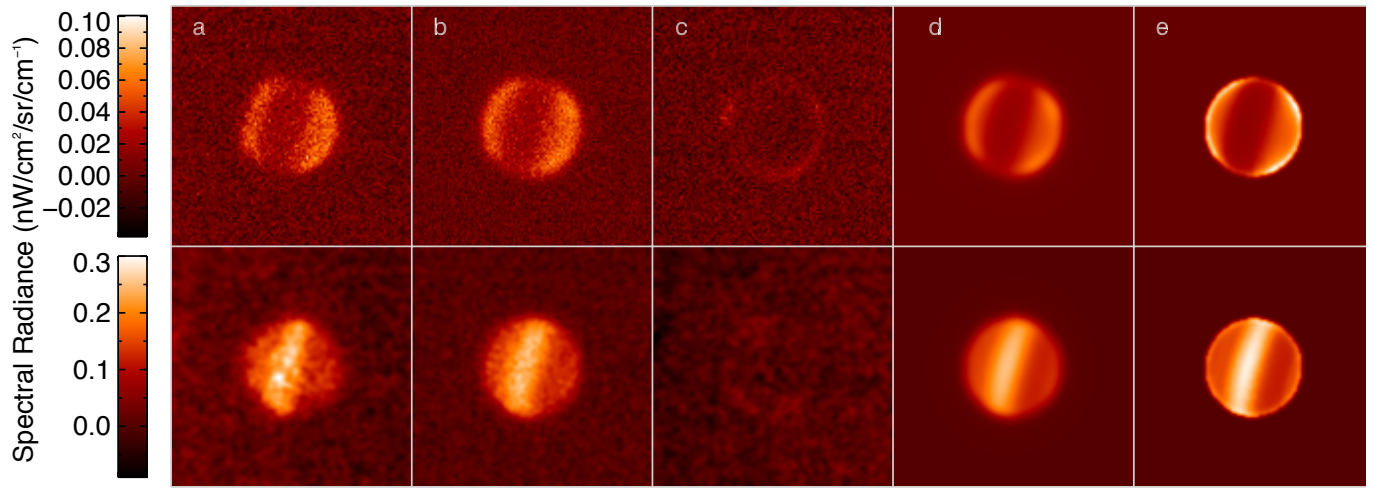}
\caption{Comparison of 2009 data and models, as in Fig \ref{fig:imgmod2018}, but for near-equinoctial data with emission at 19.5 $\mu$m (Q3, bottom) in place of the 18.7-$\mu$m data \label{fig:imgmod2009}}
\end{figure*}

\begin{figure}[ht!]
\includegraphics[width=\columnwidth, trim=0in 0in 0.0in .2in, clip]{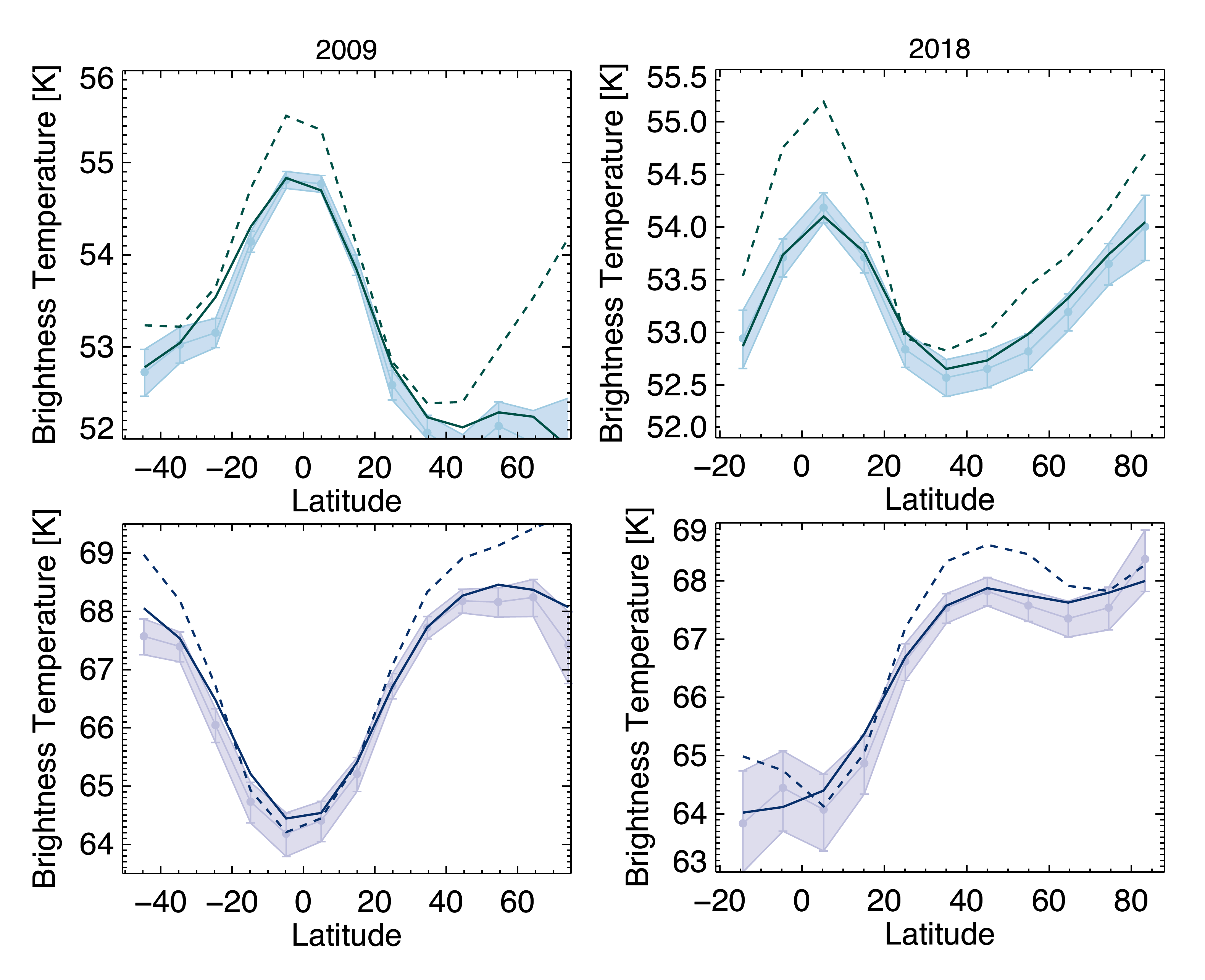}
\caption{Comparison of models to the data from which they were derived. Curves depict zonally-averaged brightness temperatures, binned 10$^{\circ}$ in latitude, extracted from both observed and modeled images. The top two plots are for the tropospheric sensing filters:  Q3 (19.5 $\mu$m) in 2009 (left) and Q2 (18.7 $\mu$m) in 2018 (right). The bottom two plots are for the stratospheric sensing filter, NeII$\_$2 (13.0 $\mu$m) in 2009 (left) and 2018 (right). The shaded regions represent the data with 3-$\sigma$ uncertainties; the darker solid line represents the model convolved with a PSF to mimic the blurring suffered by the observations; and the darker dot-dashed line depicts the brightness temperatures expected prior to convolution, representing an idealized model of intrinsic emission consistent with the observations. 
\label{fig:datvmod}}
\end{figure}

All retrievals were performed on the contemporaneous image pairs using an optimal estimation retrieval algorithm, NEMESIS \citep{irwin2008nemesis}. Calculations used collision-induced opacity based on \cite{fletcher2018hydrogen}, C$_2$H$_2$ and other hydrocarbon line data from the GEISA-2003 compilation \citep{jacquinet20112009}, and pre-computed k-distributions based on the VISIR filter properties. Additional details of the retrieval and forward modeling process are included in Appendix \ref{appendix:b}.  

With the retrieved parameters, we updated the prior 1-D model (as function of pressure) and created a 2-D model (as function of pressure and latitude). For latitudes beyond the observed domain, values were simply extended from the edge of the domain.  The retrieved atmospheric models were forward-modeled at different viewing geometries using NEMESIS and the same spectroscopic data as used for the retrievals.  

To validate our modeling approach, synthetic images using unchanged viewing geometries were also created; these forward-modeled emission were convolved with appropriate PSFs, degraded with synthetic noise, and compared to the data.  The added synthetic noise was modeled as an array of pseudorandom numbers from a normal (Gaussian) distribution with a standard deviation equal to the standard deviation of the sky in the images.  As Figures \ref{fig:imgmod2018} and \ref{fig:imgmod2009} show, the synthetic images created from the forward models simulate the data very accurately, providing confidence in our characterization of the radiance, PSF, and random noise.  The only significant deviation is near the limb in modeled 13.0-$\mu$m images, which appears only marginally brighter than the data. This likely suggests that the upper stratospheric temperatures are increasing too rapidly with increasing height in the adopted temperature model \citep{orton2014a} at altitudes above 0.1 mbar. Zonal averages derived from the data and models were also compared to demonstrate the validity of the modelled emission and the effects atmospheric seeing has on the inferred brightness temperatures (see Figure \ref{fig:datvmod}). 

In summary, models of atmospheric temperatures and acetylene were constructed from retrievals that accounted for the specific viewing geometry. These models were used to produce synthetic images that simulated the original data at alternative viewing geometries, enabling us to compare thermal emission from the different epochs.

\section{Results} \label{sec:results}
\subsection{Upper Tropospheric Emission and Temperatures: 2018 vs 1986}
\begin{figure}[ht!]
\includegraphics[width=\columnwidth, trim=.1in 0in 0in 0in, clip]{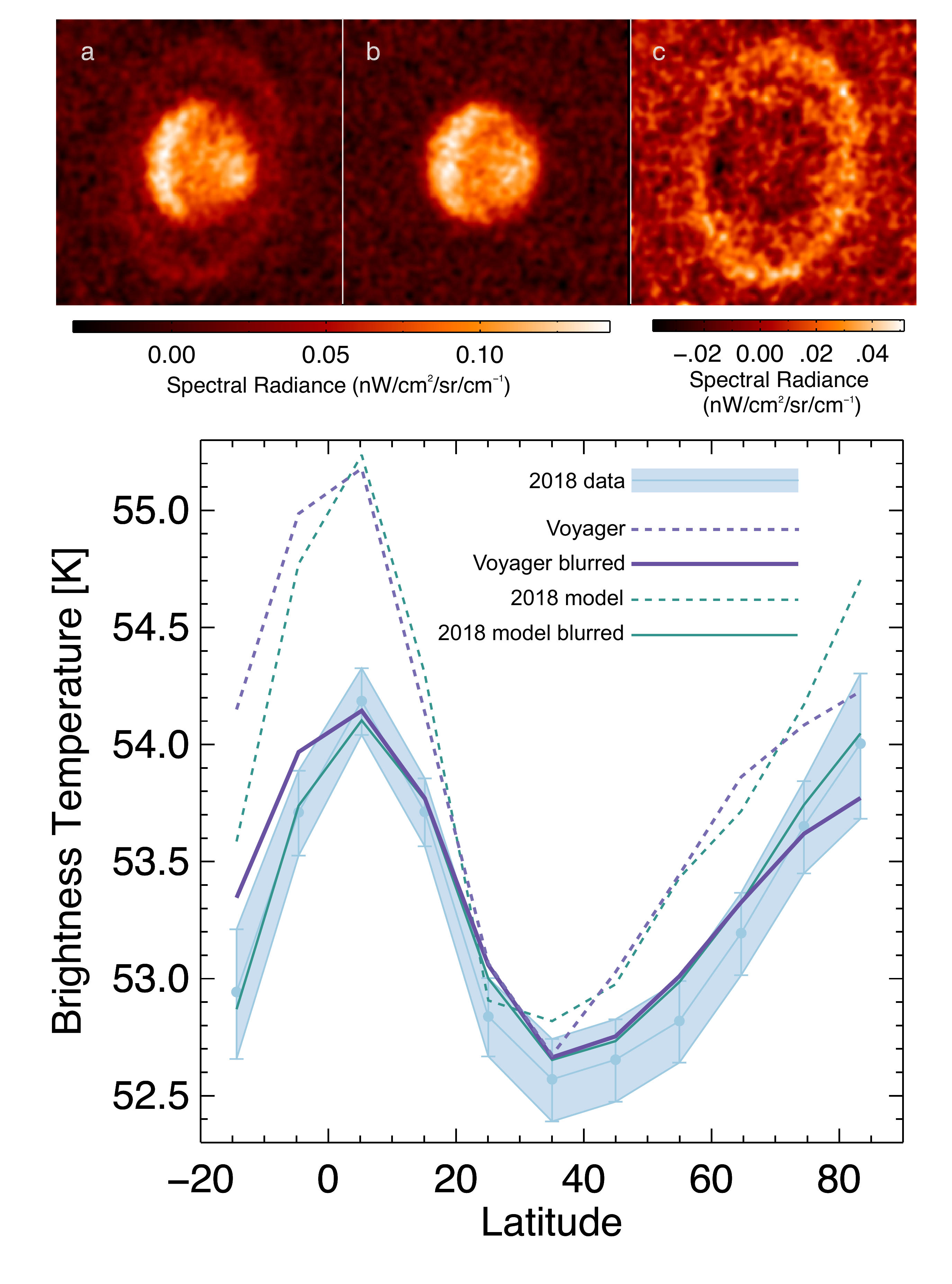}
\caption{Comparison of emission in 2018 imaging data and 1986 Voyager data. a) The 2018 18.7-$\mu$m image; b) a synthetic image created by forward modeling the temperatures derived from the 1986 Voyager IRIS spectra \citep{hanel1986infrared,orton2015thermal} to the same viewing geometry as the 2018 image.  The model was convolved with a PSF and synthetic noise was added for comparison. c) The 2018 image minus the Voyager synthetic image with slight residuals limited to the northern pole, limb, and rings (which were not modeled in our synthetic images). d) Zonally averaged brightness temperatures from the 18-$\mu$m image (shaded) from 2018 compared to the equivalent brightness temperatures computed from the Voyager data. The light shading represents the statistical 3-sigma uncertainty in the measurement centered around the mean value. The darker solid purple line shows the result of the forward-modeled Voyager emission convolved with the PSF for comparison to the ground-based data,  while the dot-dashed line depicts the Voyager brightness temperatures prior to being convolved with the PSF.  Similarly, the modeled emission derived from the 2018 data (see Figure \ref{fig:imgmod2018},d,e) was also convolved with the PSF for comparison, shown as the solid dark green curve, along with predicted emission prior to convolution (green dot-dashed curve). The curves are remarkably similar, though the polar region in 2018 appears slightly brighter. 
\label{fig:2018vsVoy}}
\end{figure}

We compared observations and zonally-averaged 18.7-$\mu$m brightness temperatures from 2018 and 1986, as shown in Figure \ref{fig:2018vsVoy}. The observed 18.7-$\mu$m image and the synthetic image derived from the Voyager temperatures appear remarkably similar after accounting for random errors and the effects of diffraction and atmospheric seeing.  The 2018 data appear only very slightly dimmer than the Voyager data everywhere except along the north edge of the planet. This can also be seen in the meridional plots of brightness temperatures derived from the images.  

For the meridional plots, firstly the 2018 brightness temperatures were simply extracted directly from the observed 2018 image and the synthetic image that had been convolved with a PSF to mimic the effects of diffraction and atmospheric seeing suffered by our ground-based observations. By this comparison, the brightness temperatures in 2018 and 1986 are equivalent to within 0.2 K at most latitudes\textemdash roughly equal to the level of statistical uncertainty (3$\sigma$, estimated from the pixel-to-pixel standard deviation of the background sky divided by the square root of the number of pixels in our zonally averaged bins). Only along the southern flank of the equatorial maximum do differences slightly exceed these estimated uncertainties, but this sharp gradient is particularly sensitive to the image resolution and the modeled PSF. 

To reduce potential differences due to imperfectly modeling the blurring, we additionally compared the 1986 brightness temperatures to equivalently blurred \textit{models} of the 2018 data.  These brightness temperatures were extracted from the blurred, modeled image derived from the 2018 retrieval (panel $d$ in Figure \ref{fig:imgmod2018} and the corresponding thick, solid curve in upper right panel of Figure \ref{fig:datvmod}). In this case, the 1986 and 2018 brightness temperatures are remarkably similar, with difference less than 0.1 K at most latitudes. Subtle differences remain near the southern edge of the equatorial maximum and the north pole. In either case, the north pole is just over 0.2 K warmer in the 2018 data compared to the 1986 data. Some of this polar brightening may be attributed to a contribution from the rings, which appear near the pole but are not simulated in our data; however, from modeling, we estimate that this contribution would account for less than 10\% of the observed difference. 

Retrievals from the data suggests these small differences in brightness temperatures could be produced by atmospheric gas temperature near 100 mbar changing by 0.3 $\pm$ 0.1 K. However, given the limited information on the vertical profile from a single image, slight changes in the retrieved profile at poorly constrained heights can partly offset values at the peak of the contribution function, leading to retrieval uncertainties that are comparable or larger than these changes. Even for the directly observed radiance, it is also important to note that we are comparing ground-based imaging to PSF-convolved forward-models based on retrievals from spacecraft data, and so small systematic errors may remain despite our attempts to account for the observational differences.  Considering the uncertainties, these results are consistent with changes of no more than 0.3 K in the brightness temperatures between 1986 and 2018, similar to what \cite{orton2015thermal} determined evaluating images dating from near equinox (ca. 2007), but with possible warming at the north pole at the limit of our uncertainty.  

\subsection{Stratospheric Emission: 2018 vs 2009}
\begin{figure*}[ht!]
\includegraphics[width=\linewidth, trim=.0in 0in .4in .0in, clip]{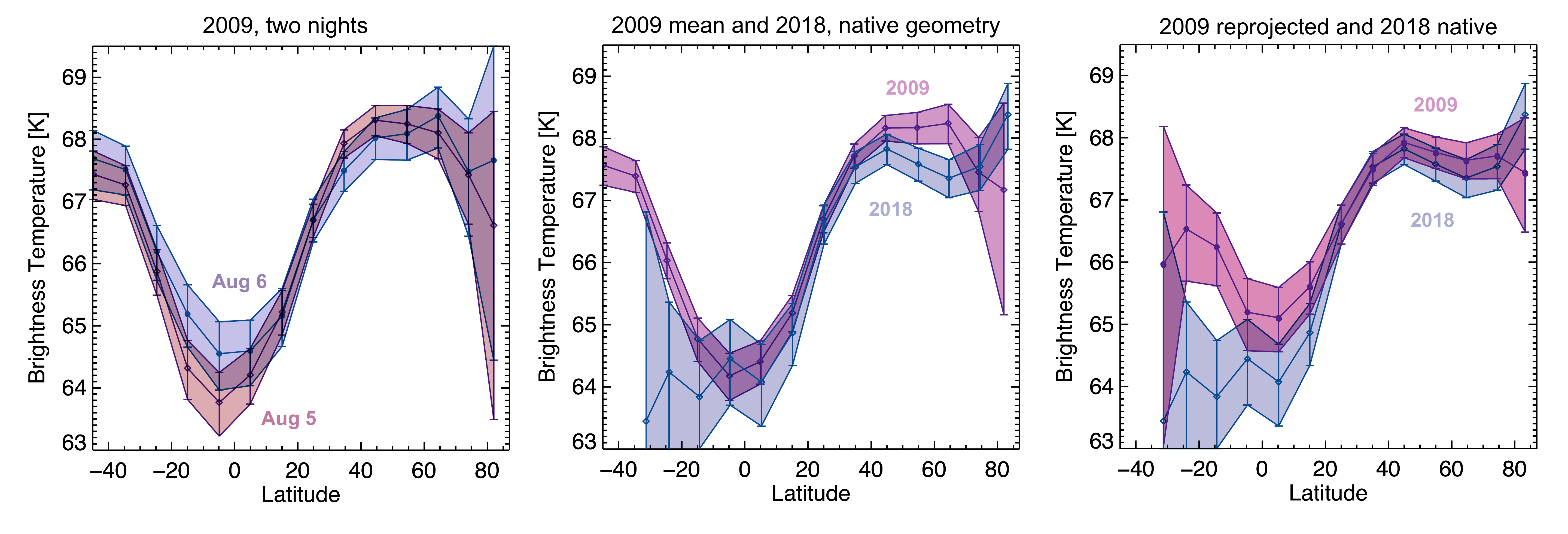}
\caption{ Zonally-averaged, binned brightness temperatures versus latitude for the 13-$\mu$m data.  Solid central curves are the mean values while shaded envelopes represent the 3-$\sigma$ uncertainties. Overlapping uncertainties are shaded more darkly and represent points where the data are statistically consistent. (Left) A comparison of the 2009 nightly averages for August 5th and 6th. The brighter equator on August 6th is just barely consistent with zonal homogeneity given the size of the uncertainty. The two nights were averaged together for our analysis and the following plots. (Middle) The mean 2009 (purple) and 2018 (blue) 13.0-$\mu$ brightness temperatures with no corrections made for different viewing geometries. The data appear mostly consistent, within the uncertainties, except for a discrepancy at higher latitudes. (Right) The atmospheric model of 2009 of data (purple), now forward-modeled to have the same viewing geometry as the 2018 data in an attempt to correct for the previous differences in emission angle, plotted over the 2018 data (blue). The model derived from 2009 data is now consistent with the 2018 data at high latitudes, but it is inconsistent in the southern hemisphere.  This indicates that either the southern hemisphere in 2018 has lower brightness temperatures than expected if 2009 conditions persisted, or that our modelling is strongly over-calculating the radiance towards the dimmer southern limb (see Fig \ref{fig:southern_change}). 
\label{fig:neii_changes}}
\end{figure*}

A comparison of zonally-averaged brightness temperatures at 13 $\mu$m from 2018 and 2009 images revealed a persistent peak in emission at northern mid-latitudes with a minimum near the equator (Figures \ref{fig:datvmod}, \ref{fig:neii_changes}). The average 2009 distribution appears to be asymmetric with a minimum just south of the equator and northern mid-latitude peak in brightness temperature that is roughly 1.0 $\pm$ 0.2 K warmer than the corresponding peak in the south (see Figure \ref{fig:neii_changes}). We note that the 2009 data showed zonal-mean variations in brightness temperatures of less than 0.5 K between the two consecutive nights, except at the equator where differences approach 1 K (Figure \ref{fig:neii_changes}, left panel). These inter-nightly differences are mostly consistent with the level of random noise in the images, but they may hint at coherent longitudinal variation (see Appendix \ref{appendix:c} for more discussion on zonal variability).

\begin{figure}[ht!]
\includegraphics[width=\columnwidth, trim=.0in 0in .0in .0in, clip]{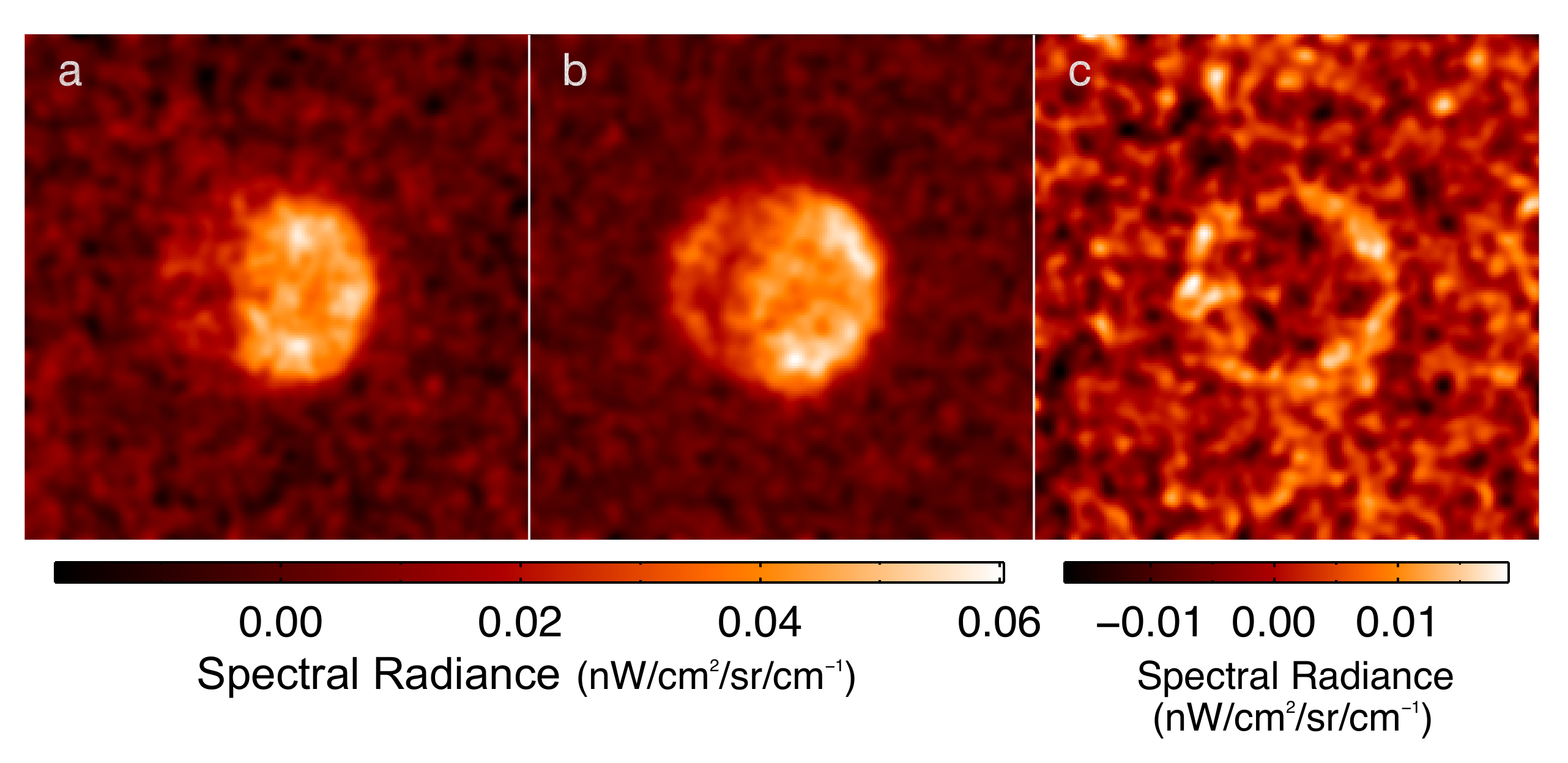}
\caption{Comparison of observed and modeled emission at 13 $\mu$m assuming the 2009 radiances persisted. a) The 2018 image showing a dark southern limb to the left in the image. b) The modeled emission assuming that the atmospheric temperatures or acetylene in 2009 had persisted until 2018.  The synthetic image of expected emission in 2018 was produced by forward modeling the retrieved temperature field from the 2009 13-$\mu$m data, though forward models created using the retrieved acetylene abundances (\textit{i.e.}, with temperatures held constant) appear identical. The forward modeled image was convolved with a PSF and corrupted with synthetic noise for comparison. c) The model minus the data shows that the model is consistently too bright along the limb, suggesting that much of the apparent brightening in the south may be due to an error in the temperature lapse rate assumed at the lowest pressures in our model and/or errors in the assumed PSF used to model atmospheric blurring. 
\label{fig:southern_change}}
\end{figure}

A comparison between 2009 and 2018 brightness temperatures extracted directly from the data at overlapping latitudes (up to emission angles of 78.5$^{\circ}$) show agreement near the center of the disk, but display discrepancies at higher latitudes. Our modeling shows that these discrepancies can largely be explained by differences in the viewing geometries (see Figure \ref{fig:neii_changes}, middle versus right panel). We showed this by essentially remapping our 2009 temperature model onto the 2018 viewing geometry (such that matching latitudes have identical emission angles) and forward modeling the expected emission (see Figure \ref{fig:southern_change}). Zonally averaged brightness temperatures drawn from this forward model show no significant change in brightness temperatures between 20$^{\circ}$ and approximately 74$^{\circ}$ north latitude. North of these latitudes, the temperature and acetylene were not actually retrieved in 2009 because of close proximity to the limb, and therefore our model of the 2009 atmosphere simply assumes an extension of the retrieved temperatures and acetylene at $\sim$74$^{\circ}$.  With this assumed atmospheric structure north of $\sim$74$^{\circ}$, the forward-modeled brightness temperatures are colder but still roughly consistent with the 2018 observations.  However, south of 20$^{\circ}$N, we find that our modeled brightness temperatures are significantly greater than the observations.  Due to the degeneracy in our retrievals, we find the same result regardless of whether we use a model of variable temperatures our acetylene in our 2009 retrievals. 

If the southern hemisphere brightness temperatures from 2009 had persisted, our modeling indicates that we should expect to see a brighter southern limb in 2018, with blurring contributing to a brighter equator. As Figure \ref{fig:neii_changes} shows, this southern emission appears absent in the data. However, as Figure \ref{fig:southern_change} shows, the difference between data and model demonstrates that this discrepancy exists along the entire limb and likely indicates a failure of the forward modeling of the limb rather than a physical change limited to the southern hemisphere. The modeled brightness along the limb is particularly sensitive to the assumed PSF (which convolves the disk and sky) and the vertical gradient in temperatures and acetylene, and so errors in either can produce modeling discrepancies.  As we noted earlier, this can indicate that the upper stratospheric temperatures are increasing too rapidly in the adopted temperature model \citep{orton2014a} at pressures less than above 0.1 mbar. Temperatures (and acetylene) at these low pressures are not retrieved in our model since the contribution functions peak deeper in the atmosphere where the acetylene abundance is expected to be greater. Lower pressures are only sensed at the very edge of the disk, where even small amounts of uncorrected blurring between the planet and background sky can suppress observed radiances and lead to erroneously lower retrieved temperatures; hence, these location (corresponding to emission angles beyond $\sim$ 72.5$^{\circ}$, or $\mu$ $<$ 0.3) were not included in our retrievals. Further investigations of the thermal center-to-limb behaviors and temperature lapse rates are beyond the scope of this paper but should be a goal of future work. 

\subsection{Temperatures and Acetylene Derived from Stratospheric Emission}\label{sec:stratTs}

\begin{figure}[ht!]
\includegraphics[width=\columnwidth, trim=.1in 0in 0.2in .0in, clip]{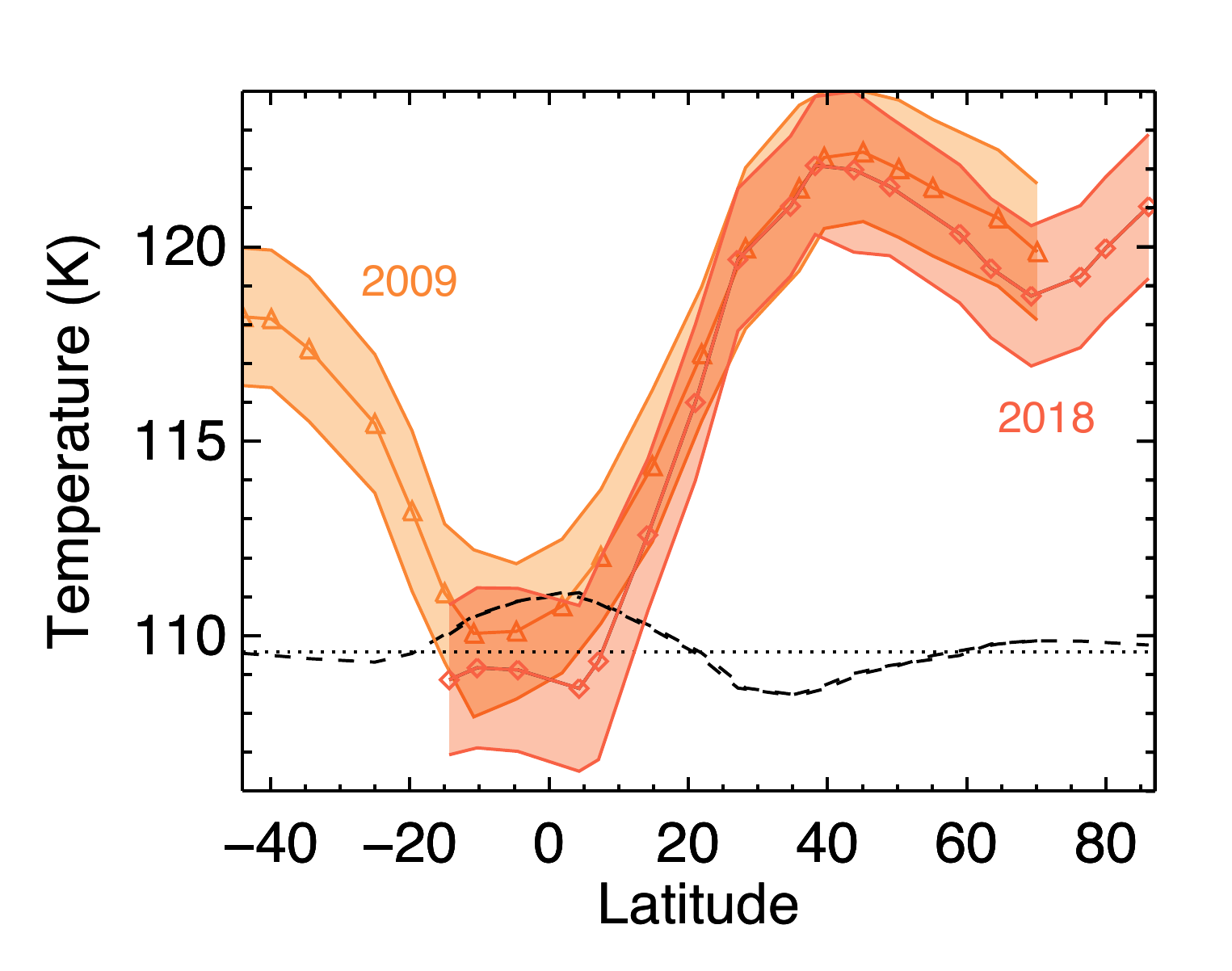}
\caption{Retrieved atmospheric temperatures at 0.25 mbar\textemdash roughly the pressure of the peak contribution at 13 $\mu$m\textemdash consistent with the meridional variability in 13.0-$\mu$m stratospheric emission assuming all variation is due to temperature change. The retrieved 2009 temperatures, shown with 3-$\sigma$ uncertainties (triangles, lighter shading), are roughly consistent with the 2018 temperatures (squares, darker shading) given the uncertainties in the retrievals. The darkest shading indicates overlap between the two curves within the uncertainties. The dotted horizontal marks the initial condition for the retrieval, while the dashed lines indicate the 0.25- mbar temperatures from the model of \cite{orton2015thermal}, which extends latitudinally resolved Voyager data to pressures less than 70 mbar by smoothly interpolating to the global 1-D temperature profile of \cite{orton2014a}. 
\label{fig:duetotts}}
\end{figure}

\begin{figure}[ht!]
\includegraphics[width=\columnwidth, trim=.2in 0in .6in .0in, clip]{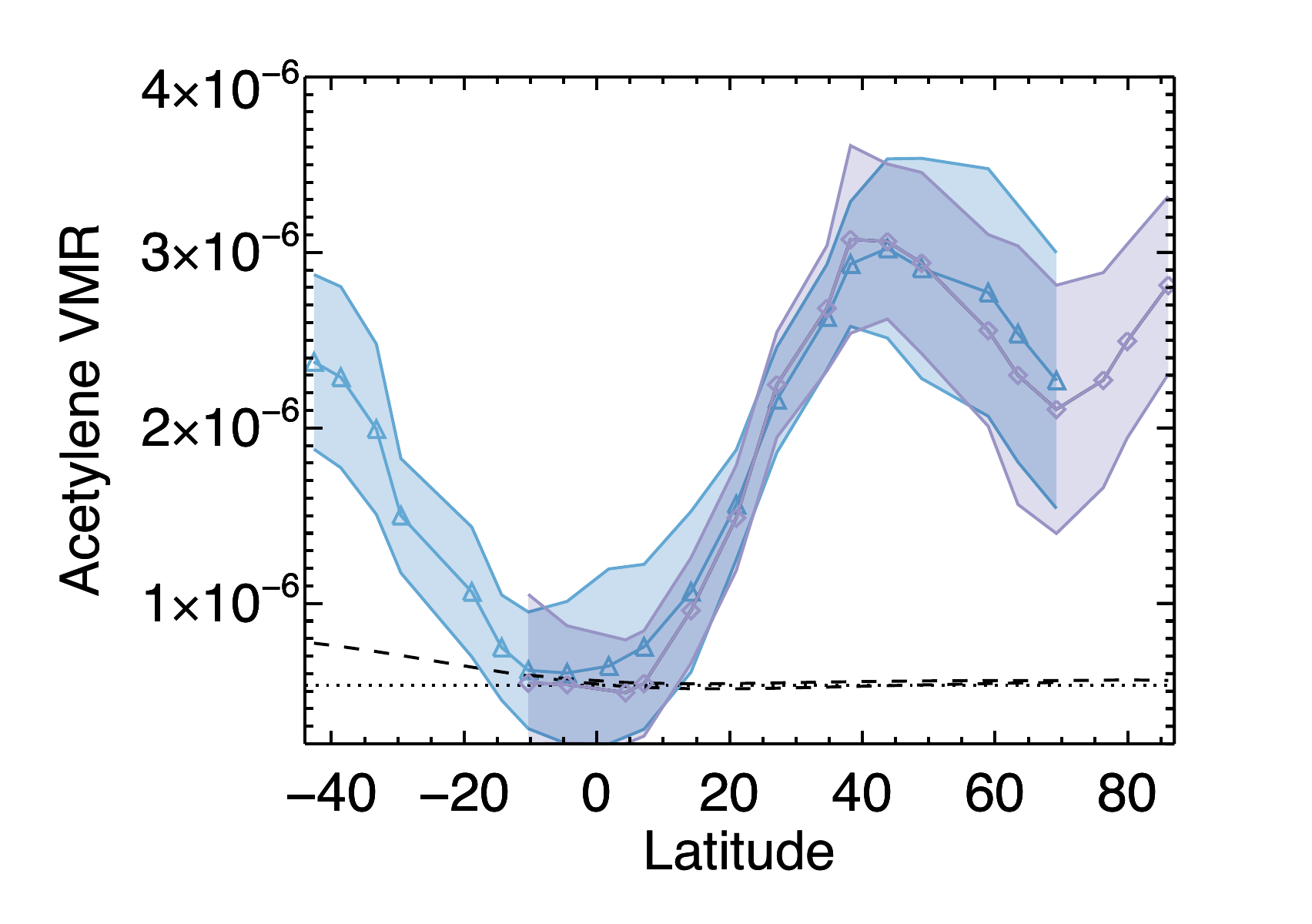}
\caption{Retrieved acetylene volume mixing ratios (VMR) at 0.25 mbar consistent with the meridional variability of the stratospheric emission (13.0 $\mu$m) assuming all variation is due to change in the acetylene abundance. The shading indicates 3-$\sigma$ uncertainties in the VMR for data from 2009 (triangles, blue shading) and 2018 (squares, purple shading).  The darkest shading indicates an overlap and a consistency between the two curves within the uncertainties. The dotted horizontal marks the initial condition for the retrieval, while the dashed lines indicate the 0.25-mbar acetylene mole fraction from the model of \cite{moses2018seasonal} for 2009 (L$_s$$\sim$7) and 2018 (L$_s$$\sim$46), nearly overlapping at mutual latitudes.
\label{fig:duetoc2h2}}
\end{figure}

To investigate sources of the stratospheric emission, we performed retrievals of temperatures and acetylene from the data. If stratospheric temperatures were allowed to vary (while acetylene was held fixed), we found the observed meridional pattern can be reproduced by a temperature gradient of roughly 13 K (13.4 $\pm$ 2.8 K in 2018 and 12.4 $\pm$ 2.8 K in 2009), measured at 0.25 mbar\textemdash roughly the pressure at which the 13-$\mu$m contribution function peaks\textemdash from the near-equatorial minimum to the northern mid-latitude peak at $\sim$40$^\circ$ (see Figure \ref{fig:duetotts}). In the southern hemisphere, the peak temperature was roughly 4 K less.  The variation in radiances were equally reproduced by a 500-600\% increase in the acetylene mixing ratio at northern mid-latitudes ((2.7 $\pm$ 0.7)$\times$10$^{-6}$ VMR increase in 2018 and (2.0 $\pm$ 0.6)$\times$10$^{-6}$ increase in 2009) relative to the expected values of \cite{moses2018seasonal} and those retrieved at the equator. There was roughly 20\% less at southern mid-latitudes compared to the north, but these southern latitudes are only sampled in 2009 nearer the edge of the disk (see Figure \ref{fig:duetoc2h2}).  

While an increase in stratospheric acetylene had no effect on the inferred underlying upper tropospheric temperatures, we found that stratospheric warming would also be detectable at 18.7 $\mu$m because of the broad contribution function of the Q2 filter. Therefore, we find that warming in the stratosphere must be compensated by cooling the troposphere to remain consistent with the observations. If the 13.0-$\mu$m emission is due to increased stratospheric temperatures, then the inferred tropospheric temperatures may be up to 1 K colder at the mid-latitudes relative to those inferred from the Voyager-IRIS data (assuming the longer wavelength Voyager measurements were not equally influenced by the contribution of warmer stratospheric temperatures).  This difference is nearly comparable to the potential uncertainty in retrieved temperature profiles, but it would be consistent with cooling of the northern (autumnal) hemisphere following equinox. 

\section{Discussion} \label{sec:discussion}
\subsection{Seasonal Changes}

The 1986 Voyager data reveal a meridional cross section of Uranus' temperatures near the tropopause shortly after the southern summer solstice. If we assume the atmospheric temperatures respond to solar heating and the radiative relaxation times approached a full season, the 1986 Voyager data would have displayed the atmosphere's thermal response to solar forcing at the time of the preceding equinox in 1966 \citep{conrath1990temperature}.  It then follows that the equinoctial data analyzed by \cite{orton2015thermal} would have provided a view of both hemispheres responding to asymmetric solstitial (southern summer, northern winter) solar forcing (although little seasonal asymmetry was detected), and our 2018 observations would show the northern hemisphere's transition from winter to spring. The northern mid-latitudes remained in darkness for nearly a decade past solstice in 1986, so it is plausible that cooling may have progressed beyond the 2007 equinox at these latitudes if the seasonal response lagged a full season behind the solar forcing.     

Though uncertainties are significant, our data show that changes near the tropopause are consistent with atmospheric temperature changes no greater than $\sim$0.3 K since solstice, though perhaps slightly larger if potentially elevated stratospheric temperatures contributed to our retrieved temperatures (see Sec \ref{sec:stratTs}). These values are consistent with the conclusion of \cite{orton2015thermal}, who reported brightness temperature changes of less than 0.4 K between equinox and solstice.   

The seasonal amplitude expected from radiative forcing will in general be reduced by a factor of $1/\sqrt{1+r^2}$ where $r=2\pi \tau_R/\tau_O$ and $\tau_R/\tau_O$ is the ratio of the radiative time constant to the orbital period \citep{conrath1998thermal}. \cite{conrath1990temperature} showed that seasonal forcing of 10 K and a radiative time constant of 130 years should result in changes of 1 K or less.  Using the updated, shorter time constants of \cite{li2018high}, this value would be as large as 2\textendash7 K with sub-seasonal lags (although this assumes the same 10 K forcing which may not be consistent with the radiative heating applied by the authors). The amplitude and phasing derived from our data favor timescales of over 300 years, which, considering the seasonal scales, amounts to essentially no seasonal variation due to solar forcing at mid-latitudes.  We emphasize that this is again based on seasonal radiative equilibrium temperatures from \cite{conrath1990temperature}.  Shorter radiative timescales would suffice if different heating rates and dynamical redistribution are assumed.  \cite{bezard1986model} computed seasonal contrasts of 4 K or less in the radiative equilibrium temperatures (depending on the latitude dependence of the internal heat flux), which would be consistent with radiative cooling times on order of a century. Shorter timescales computed by \cite{li2018high} would require equilibrium temperature swings of under 1.5 K or simply a greater amount of meridional mixing to overwhelm the radiative response \citep{friedson1987seasonal}.  An updated radiative-dynamical model using latest absorption coefficients and chemical abundances is needed to better evaluate the consistency of these radiative time constants with the observations.

Higher in the stratosphere, our comparison is limited to a span of nine years. This span is short compared to estimates of long radiative and dynamical timescales on Uranus \citep{conrath1990temperature,moses2018seasonal}, and so it is unsurprising to find little change in the northern hemispheric emission at 13 $\mu$m. By the same token, it would be surprising if any physical process could produce the the apparent discrepancies between the 2018 data and 2009 model, further suggesting that this is likely attributable to an error in the assumed profiles of temperatures or acetylene abundances at higher altitudes that go into producing the forward model.

\subsection{Nature of the 13-$\mu$m Emission}
\begin{figure}[ht!]
\includegraphics[width=\columnwidth, trim=.1in 0in .0in .0in, clip]{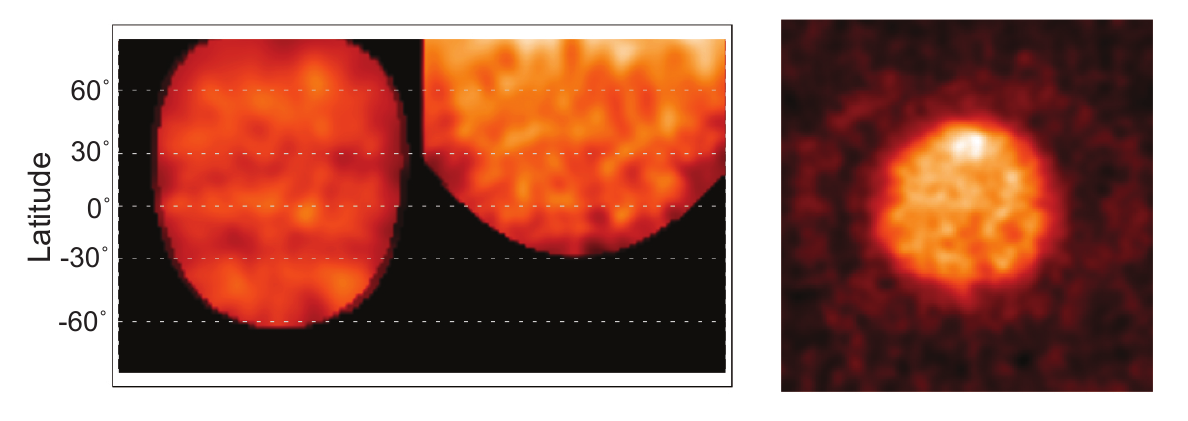}
\caption{ The combined sum of the 13-$\mu$m and scaled 18.7-$\mu$m maps and images, where the anomalies essentially cancel each other out, illustrating the anti-correlation of radiances in the data at equatorial and mid-latitudes. At the pole, the anomalies are both positive, leading to an enhanced bright spot.  Together these are consistent with upwelling at mid-latitudes and downwelling at the pole. Radiances at 18.7 $\mu$m were reduced by 50\%. 
\label{fig:inverse}}
\end{figure}

The cause of the 13-$\mu$m emission distribution is unknown, but as discussed in Section \ref{sec:stratTs}, it could result from a regionally elevated temperatures, acetylene or a combination of the two. Though not obvious from each pair of images alone, maps of the data show a remarkable anti-correlation between the 13-$\mu$m and 18.7-$\mu$m radiances (see Figures \ref{fig:vs_other_data} a,b and \ref{fig:inverse}). Furthermore, while the mid-latitudes are negatively-correlated, both images show brightening at the poles. Together, these correlation suggest that a shared mechanism or mechanisms may be responsible for the emission in both filters, forming a dynamical link between stratosphere and upper troposphere.

One possible explanation for the observed emission may be elevated temperatures resulting from the adiabatic compression of regionally subsiding gas.  Just as the meridional temperature structure in the upper troposphere has been interpreted as a consequence of large-scale vertical motions \citep{flasar1987voyager}, similar but separate circulation cells could possibly be at work higher in the stratosphere, rotating in an opposite sense to produce downwelling at mid-latitudes and upwelling at the equator. The meridional position of this stacked circulation would not be merely coincidental, as it could be linked to the underlying meridional temperature gradient.  While the driving force behind the upper-tropospheric circulation is unknown \citep{flasar1987voyager}, the meridional temperature structure it produces has a maximum temperature gradient between 15$^{\circ}$ and 20$^{\circ}$ latitude in both hemispheres (with dT/dy$<$0).  It is conceivable that this upper tropospheric temperature gradient could be geostrophically balanced by an unobserved stratospheric tropical jet aloft via the thermal wind relationship. Potential dissipative processes weakening the thermal wind with height could then lead to a mass-balancing meridional circulation with descending, warming air on the poleward side of the jet \citep{conrath1983thermal}. Following \cite{flasar1987voyager} (their Eq. 3), we can relate the desired steady-state temperature differences to a requisite differential vertical velocity by the expression 
\begin{equation}
\Delta w = \Delta T R / \tau_R H N^2 
\end{equation}\label{eq:eq1} where $\Delta T$ is the meridional temperature difference, $R$ is the gas constant ($\sim$ 3614.9 J/kg/K), $H$ is the pressure scale height ($\sim$ 47 km), $N$ is the buoyancy frequency (0.0048 $s^-1$), $\tau_R$ is the radiative time constant, and ($\Delta w$) is the vertical velocity difference.  Assuming $\tau_R$ $\sim$ 49 yrs at 0.2 mbar \citep{li2018high} and $\Delta T$ $\sim$ 13 K from our retrievals, we calculate $\Delta w$ $\sim$ 2.8$\times$ 10$^{-5}$ m/sec\textemdash five times greater in magnitude than what \cite{flasar1987voyager} computed for the tropospheric vertical velocity differential.

While adiabatic warming associated with thermal winds could potentially explain the enhanced emission at mid-latitudes, a different mechanism would be needed to explain the pattern in emission at higher latitudes.  The apparent second peak in emission at the pole would be inconsistent with with a thermal wind-driven circulation since the tropospheric temperature gradient is in the opposing direction at high latitudes (\textit{i.e.}, dt/dy$>$0.) A separate mechanism such as wave heating or a larger, independent circulation would be needed to heat the pole. Ultimately, assessing the dynamical feasibility of this scheme requires evaluating the thermal wind equation and corresponding mass conservation for zonal, meridional, and vertical components of the winds, which is beyond the scope of the present data and this paper. 


Alternatively, a regional enhancement in acetylene could satisfy the 13 $\mu$m emission with a much simpler and coherent dynamical circulation. If the putative mid-latitude upper tropospheric upwelling simply extended further into the stratosphere ($\sim$6 scale heights above the tropopause), tropospheric methane could potentially be mixed to lower pressures where it would photolyse to produce acetylene. As long as chemical conversion timescales were short compared to timescales for meridional transport, the concentration of acetylene would peak in the region of upwelling and diminish quickly to the north and south. \cite{moses2018seasonal} suggests acetylene loss timescales are roughly 2 years, with net lifetimes of 40 years at 0.2 mbar. Timescales of meridional transport are uncertain but potentially much larger, with \cite{conrath1998thermal} suggesting a dynamical timescale of 700 years in the stratosphere. 

The resulting enhancement would be consistent with inferences of latitudinal variation in stratospheric hydrocarbons from the Voyager UVS data \citep{yelle1989far,mcmillan1992dynamical}. \cite{yelle1989far} reported a factor of 2 to 3 reduction in the observed reflectance in the 1338-1583 \AA{} spectral band at mid-latitudes relative to the pole and interpreted this signal as relative depletion of hydrocarbons at the poles and enrichment due to extended upwelling at lower latitudes. For consistency, it then follows that the brightness detected at the poles at 13 $\mu$m and 18.7 $\mu$m could both be potentially attributed to adiabatic warming in the sinking branch of this extended cell. 

Regional enhancement of acetylene would also imply that methane would be similarly enhanced at mid-latitudes. Though not detected in near-IR data \citep{karkoschka2009haze}, regional enhancements in methane are indirectly consistent with inferences of supersaturated and spatial variable methane from \textit{Herschel PACS} spectra \citep{lellouch2015new}, which found supersaturated methane mole fractions of $\sim$9.2 $\times$ 10$^{-5}$ at the tropopause\textemdash roughly six times larger than the value inferred from Spitzer \citep{orton2014mid}. The authors noted that the measurements could be reconciled if the methane profile decreased by a factor of five from 100 to 2 mbar, but such a profile would be difficult to explain. They argued that the discrepant supersaturated mole fraction may have instead indicated spatial heterogeneities in stratospheric methane abundance or temperatures, noting that the \textit{Hershchel} data probed more global conditions while the \textit{Spitzer} data were biased towards warmer temperature regions.  Interestingly, our data indicate a factor of 5\textendash6 enrichment in acetylene in precisely the coldest regions (i.e., mid-latitudes) while mole fraction at warm equatorial regions remain in strong agreement with values derived from \textit{Spitzer} \citep{orton2014mid, moses2018seasonal}, consistent with proposed explanation for the observed discrepancy. 

A vertically extended circulation cell has already been suggested for Neptune to explain correlations between observations in the stratosphere and troposphere \citep{de2014neptune}, and was previously proposed for Uranus by \cite{yelle1989far} and \cite{mcmillan1992dynamical}. It is perhaps unsurprising that a similarly extended circulation may be found in the stratosphere on both planets, given their qualitatively similar, broad tropopauses \citep{de2014neptune}, despite the weaker vertical mixing and stronger seasonal forcing on Uranus.  In either case, the upward flux of methane may have to exceed the presumed volume mixing ratio limit imposed by the estimated equilibrium saturation vapor pressure at the tropopause (i.e., the cold trap). Evidence of potentially supersaturated mixing ratios of methane on Uranus and especially Neptune suggest that vertical mixing may indeed overcome the cold trap limitation \citep{lellouch2015new}, though the mechanisms of this process remain speculative \citep{lunine1989abundance,de2014neptune}. Inversely, the greater vertical extent of the circulation cell would mean that upwelling gas would experience a greater and more extended flux of high energy photons along its extended path compared to a circulation cell that ceased nearer the tropopause; this would serve to further deplete the gas of chemicals destroyed by photolysis, meaning only the most stable species would be present in the downwelling branches. 

A strong coupling of the troposphere and stratosphere would also help to physically link the apparent hemispheric asymmetry seen in the tropospheric temperatures and 13-$\mu$m emission. Both Voyager and equinoctial 13-$\mu$m imaging data show the equatorial maximum/minumum to be slightly offset to the southern hemisphere \citep{orton2015thermal,orton2018neii}. Likewise, Voyager spectra and ground-based 18-$\mu$m imaging showed the mid-latitude temperature anomaly in the northern hemisphere to be roughly 1 K colder (i.e., a $\Delta$T of $\sim$4 vs 3 K), while the 13 $\mu$m emission from the northern hemisphere in 2009 appears roughly 1 K brighter than the southern hemisphere (a retrieved acetylene VMR of 3.0$\times$10$^{-6}$ vs 2.3$\times$10$^{-6}$). Both of these represent a roughly 30$\%$ difference that could be explained by upwelling that is greater in the north than in the south. Considering this asymmetry and the potential dynamical link, the apparent absence of emission along the southern limb in 2018 data may be an indication of an asymmetric extent of the upwelling, such that the acetylene mixing ratio and the contribution function peak at greater pressures and hence become less visible in the extended optical paths along the limb. Or perhaps the strength or position of the upwelling has changed in time (e.g. a slight southward shift of the southern hemisphere upwelling) due to deeper dynamical processes. Unfortunately, the southern hemisphere will soon be unobservable from Earth for several decades, so testing theories and confirming changes will have to remain a goal of future work and potentially visiting spacecraft. 

In contrast to the elevated temperature scheme due to downwelling, upwelling would actually serve to reduce the temperature at mid-latitudes through adiabatic cooling. Therefore, the alternative explanations of stratospheric downwelling and hydrocarbon upwelling would serve to produce temperature anomalies in opposite directions, although the cooling would be relatively minor due to the much weaker updrafts expected at these heights \citep{flasar1987voyager}. Independent observational constraints on the meridional temperature structure in the stratosphere can come from sensing the H$_2$ S(1) quadrapole line and thus could help discriminate between these two theories.  \cite{trafton2012mid} note that 2007 Gemini TEXES observations show a bi-modal but asymmetric pattern that is brightest towards the north, but further work is needed to determine the consistency with the above models. 

One potential flaw with the extended upwelling mechanism is the apparent inconsistency at the pole and the equator.  If downwelling produces adiabatic heating in the upper troposphere (i.e., at 18 $\mu$m) at the poles and equator, why does stratospheric downwelling appears to heat the stratospheric pole but not the equator?  Even if the effect of the adiabatic heating was offset by the reduced acetylene, it is not obvious why the equator and pole should be different. Considering mass conservation, the considerably lesser surface area of the polar regions relative to the equatorial regions may naturally be a factor in producing greater adiabatic heating at the pole, but then we would also expect a similar effect in the deeper 18.7-$\mu$m emission unless the area of subsidence expands with increasing depth. One possible solution is that the pole is also heated by independent dynamical mechanisms such as breaking waves or greater concentrations of absorbing aerosols, but this remains purely speculative. More detailed dynamical modeling will be necessary to help evaluate the circulation in the stratosphere, its link to the troposphere, and its overall changes in time. 

\subsection{Comparison to Visual/Near-IR Imaging }
\begin{figure*}[ht!]
\includegraphics[width=\linewidth, trim=.0in 0in .0in .0in, clip]{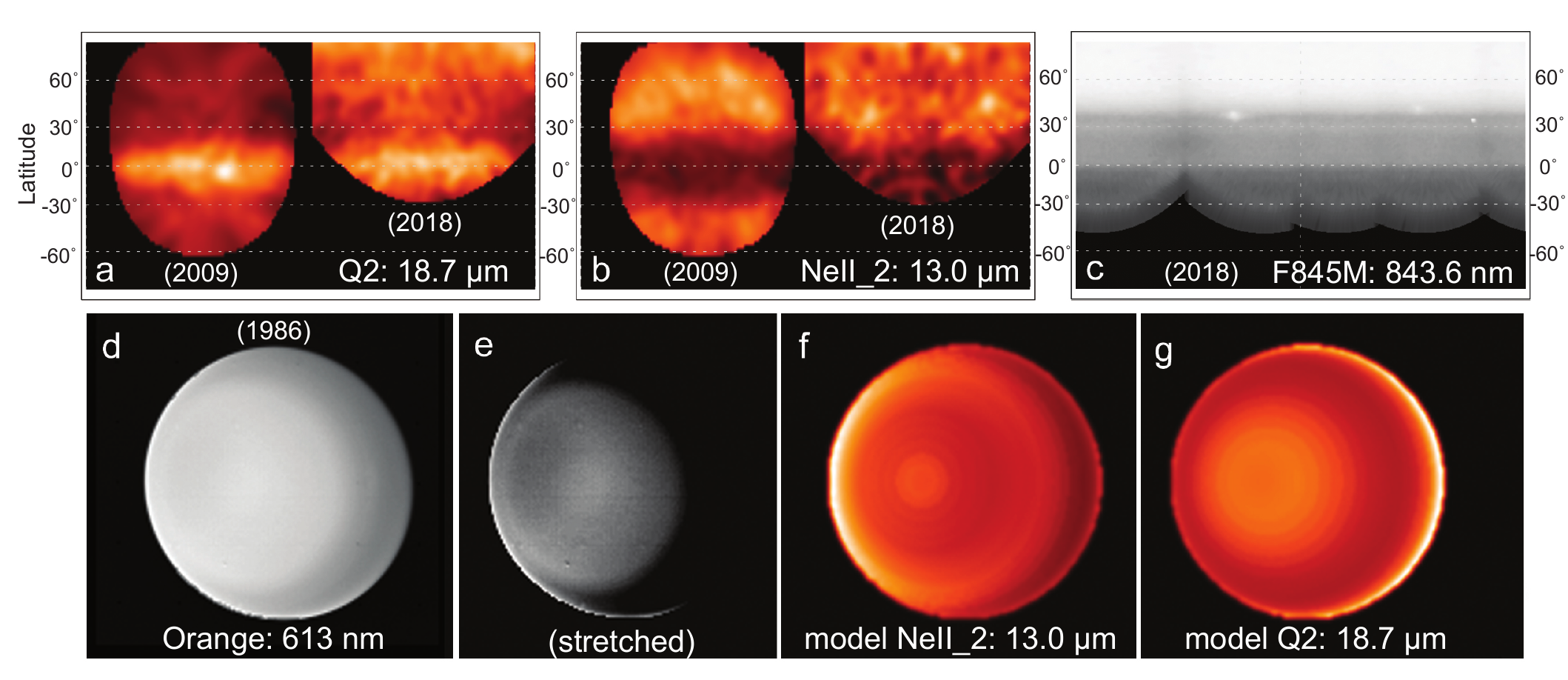}
\caption{Comparison of the thermal data and models to deeper near infrared and visible imaging. a) Mapped 19.5-$\mu$m (Q3) and 18.7-$\mu$m (Q2) VLT-VISIR images showing emission from the upper troposphere, with a peak at the equator and a minimum at mid-latitudes. b) Mapped 13.0-$\mu$m (NeII$\_$2) VLT-VISIR images showing stratospheric emission peaking at mid-latitudes and a clear minimum at the equator. c) Hubble Space Telescope (HST) WFC3/UVIS map mosaic of in the near-IR (F845M filter) from November 16, 2018 ($\sim$1 month following our VLT observations). The prominent seasonal polar haze is seen above Uranus with a boundary around $\sim$40$^{\circ}$, straddled by two small discrete features, coinciding with the peak of the 13-$\mu$m emission.  The edge of the 13-$\mu$m emission extends 15$^{\circ}$ S of the polar haze boundary. d) Voyager 2 image of Uranus in the orange filter (613 nm) taken with the Narrow Angle Camera on January 4, 1986.  e) The same image is stretched to enhance the polar-bright structure, which is similar in appearance to f), the modeled radiance in the stratosphere inferred from the data, and vaguely similar to g), the warming seen in our model derived from contemporaneous Voyager-IRIS spectra.  Note that this Voyager-inferred south-polar emission extends more broadly in latitude compared to the equivalent north-polar emission retrieved from 2018 data (see Figure \ref{fig:imgmod2018}).
\label{fig:vs_other_data}}
\end{figure*}

A comparison of the mid-infrared images to visible and near-IR imaging, dominated by absorption from methane and the scattering from clouds and hazes at greater pressures, show some possible signs of correlations between the stratosphere and troposphere (Figure \ref{fig:vs_other_data}).  Although the 13-$\mu$m emission extends $\sim$15$^\circ$ further to the south, the boundary of the brighter polar northern haze layer in a near-IR HST image is at roughly the same latitude that the 13-$\mu$m emission peaks ($\sim$40$^\circ$).  Intriguingly, the bright discrete cloud features near 40$^\circ$ in the HST image also roughly correspond in latitude to the brightest discrete signals seen in the nightly averaged 13-$\mu$m images. The tenuous correlation might suggest that tropospheric vortices generate localized hydrocarbon enhancements through upwelling \citep{de2014neptune} or localized heating via breaking waves, although a thorough analysis of the image noise suggest that these radiances would constitute noise at the $\sim$2.5-$\sigma$ level and hence be inconclusive (see Appendix \ref{appendix:c}). Establishing a link between upper-tropospheric vortices and the stratosphere will require greater signal-to-noise imaging in future work. 

If we assumed hemispheric symmetry in the meridional structure of the 13-$\mu$m polar emission, we can compare our thermal model to images from Voyager. The polar aerosols seen from this perspective show an enhanced brightness very near to the pole in Voyager's orange (590\textendash640 nm) and methane (614\textendash624 nm) filters, with an extent that is similar to the pattern seen in our models derived from the stratospheric emission. The sharper increase at the pole is not evident in the Voyager-IRIS thermal spectra, which shows a more gradual warming roughly coincident with the polar haze region.  We note that we do see a similar sharper feature at the north pole in our 18.7-$\mu$m 2018 data (see Figure \ref{fig:imgmod2018}) even though it appears absent at the synthetic images generated from the Voyager spectra. If linked across the troposphere and stratosphere, these observations may indicate that stratospheric downwelling or photochemistry at the pole affects the albedo or abundance of the tropospheric haze layer $\sim$9 scale heights below. This may potentially be accomplished by methane-depleted air settling from above, combined with photochemical products produced in the higher ultraviolet flux found in the stratosphere. Assuming the parent chemicals were derived from upwelling gases, the aerosols or particulates would have to be stable enough to survive meridional transport from mid-latitudes to the pole and down to the cloud level, potentially over centuries \citep{conrath1998thermal}. 

Considering the recent detection of hydrogen sulfide above the cloud layer \citep{irwin2018detection}, a possible candidate for stratospheric aerosols may be stable sulfur allotropes, as investigated as a possible source of haze in exoplanetary atmospheres \citep{zahnle2016photolytic,gao2017sulfur}. Trace amounts of upwelling H$_2$S could photodissociate in the stratosphere and combine with free H atoms (produced from the photolysis of CH$_4$) to eventually form stable octasulfur allotropes (S8) capable of surviving further UV radiation and the range of temperatures found above the clouds. If carried with the circulation and concentrated at the pole before settling down upon the clouds, these yellow particulates \citep{meyer1972spectrum,eckert2003molecular} could potentially explain the scattering preferentially seen in the orange (614 nm) filter relative to the blue (477 nm) and violet (431 nm) filters \citep{smith1989voyager}. Though, it is worth noting that this is purely speculative and several other species of hydrocarbon polymers absorb in the violet and may potentially explain the visible spectra \citep{baines1986structure,pollack1986estimates}.

Polar aerosols could also potentially explain the slightly warmer summer temperatures at the poles in the lower stratosphere seen the Voyager data \citep{orton2015thermal}. If this was due to solar heating, such changes would only be expected from the radiative-dynamical model of \cite{conrath1990temperature} if radiative time constants were significantly shorter than even those of \cite{li2018high} at stratospheric heights. This could potentially be caused by unaccounted presence of absorbing aerosols. Likewise, the broader polar brightness seen in the model 18.7-$\mu$m emission derived from the Voyager IRIS spectra may indicate broader subsidence or additional heating associated with overlying hazes or deeper tropospheric aerosols seen in the Voyager imaging.  However, in terms of seasonal changes, the vertical velocities implied by the temperature variations ($\sim$ $10^{-5}$ m s$^{-1}$) are still comparable or small compared to the expected fall velocities of photochemical aerosols \citep{toledo2019constraints}. Therefore, if estimates of settling times and vertical velocities are correct, downwelling may not significantly aid the transport of material from the stratosphere to the cloud level on timescales short enough to explain seasonal changes in the appearance of polar regions. Therefore, any apparent correlations in structure may be signs of the mean annual signatures of transport imprinted on the seasonally varying albedo of the haze layer, itself controlled by mechanisms that have yet to be fully explained. If seasonal changes are related to temporal changes in the methane mixing ratio above the cloud layer, it is conceivable that this process may be aided or triggered by a seasonal increase in the rate of subsidence.  If true, this process may be detectable by measuring changes in the polar thermal emission, although as our data show, it is challenging to measure the polar regions near equinox given the geometry seen from Earth. 

If variations in the seasonal albedo are related to changes in temperatures\textemdash whether through changes in condensation, subsidence,  or convective stability\textemdash these temperature changes are not apparent in our upper tropospheric or stratospheric data. \cite{li2018high} computed radiative time constants as short as a decade near the polar cloud tops \citep{sromovsky2019methane}, but the variation of cloud layer temperatures have yet to be measured. We have shown that radiative time constants at higher altitudes are either longer than expected or mixing can effectively dampen seasonal temperature changes.  If  meridional mixing is present in the variable cloud layer as well, it apparently does not traverse the boundary at 40$^{\circ}$ latitude. 

\section{Conclusions}\label{sec:conclusions}
Analysis of our ground-based mid-infrared imaging of Uranus by VLT/VISIR in 2018 has revealed persistent thermal structures of the troposphere and new insight into the circulations of the stratosphere.  

Uranus' upper tropospheric temperatures have changed little since the 1986 solstice, with the basic structure consistent with persistent mid-latitude upwelling \citep{flasar1987voyager}. Brightness temperatures measured from zonally averaged meridional profiles are consistent with changes of less than 0.3 K in the 32 years since the southern summer solstice, with possible warming at the north pole due to adiabatic compression, aerosol heating, or other dynamical processes. These small changes are consistent with extremely long radiative time scales or very efficient meridional heat transport, suggesting the need for updated radiative-dynamical models. 

To evaluate potential changes in the stratosphere, we compared imaging data at 13 $\mu$m from 2018 and 2009 and found little change in the northern mid-latitudes. We find a significant asymmetry between the northern and southern mid-latitudes in 2009 images that may be associated with an asymmetry in dynamical mixing or unexpected changes in the photochemisty. Though constraining observations of the southern hemisphere may not be available again until the southern hemisphere returns into view in the late 2030s, modeling should aim to test potential explanations for the hemispheric asymmetry. 

The meridional structure of the stratospheric emission nearer the pole came into view in 2018, allowing us to infer a possible peak at mid-latitudes that appears remarkably anti-correlated with the 18 $\mu$m emission from deeper in the atmosphere.  The 13 $\mu$m mid-latitude peak also shows possible longitudinal structure that roughly coincides with the bright cloud features and a transition in reflectance within the cloud layer below.  We interpret these correlations as indicators of a potential link between the stratosphere and troposphere. Although the stratospheric emission may be due to regionally warmer temperatures produced by downwelling, we believe an enhancement of a factor of roughly five in the acetylene mole fraction at mid-latitudes can explain the observations more simply. This implies that the upper-tropospheric circulation pattern inferred from Voyager data \citep{flasar1987voyager} extends at least six scale heights into the stratosphere and is capable of transporting hydrocarbons higher than previously appreciated. Thus, we interpret the observed pattern of acetylene as possible evidence that methane is primarily transported from the troposphere, through the cold trap, and into the stratosphere where it is subsequently photolyzed into acetylene and limited to mid-latitudes. This would suggest a commensurate enhancement of methane near the tropopause at mid-latitudes, potentially consistent with supersaturated or spatially variable methane profiles inferred from \textit{Herschel} \citep{lellouch2015new}.  We argue that the long path through the stratosphere along this extended circulation cell leads to greater photolytic depletion of hydrocarbons in the corresponding downwelling branches expected at the equator and poles, accompanied by adiabatic warming that extends from the middle stratosphere to at least the upper troposphere, particularly at the poles. 

Our analysis is based on a limited number of noisy images and much of what we discussed is speculative, given uncertainties in the data and a lack of published dynamical models. The unprecedented sensitivity of the James Webb Space Telescope promises to greatly improve our understanding by providing unambiguous characterization of the thermal structure of the north pole in the 2020s.  Dynamical models of Uranus' uniquely forced atmosphere and obscure circulations will be needed to help interpret present observations and the many unexpected findings to come.

\acknowledgments
This analysis was based on observations collected at the European Organisation for Astronomical Research in the Southern Hemisphere under ESO programmes 0101.C-0073(B) and 083.C-0162(A). This work was supported by a European Research Council Consolidator Grant, under the European Union’s Horizons 2020 research and innovation programme, grant number 723890. LNF was also supported by a Royal Society Research fellowship. GSO was supported by NASA through funds distributed to the Jet Propulsion Laboratory, California Institute of Technology.  This work used the ALICE supercomputing facilities provided by the university of Leicester. This work used data acquired from the NASA/ESA HST Space Telescope, associated with OPAL program (PI: Simon, GO13937), and archived by the Space Telescope Science Institute, which is operated by the Association of Universities for Research in Astronomy, Inc., under NASA contract NAS 5-26555. All maps are available at \dataset[MAST DOI 10.17909/T9G593]{http://dx.doi.org/10.17909/T9G593}.

\bibliography{mybib} 
\bibliographystyle{apj}

\appendix

\section{Filter Transmissions and Contributions}\label{appendix:contributions}

\begin{figure}[h]
\includegraphics[width=\columnwidth, trim=.1in .1in .1in .1in, clip]{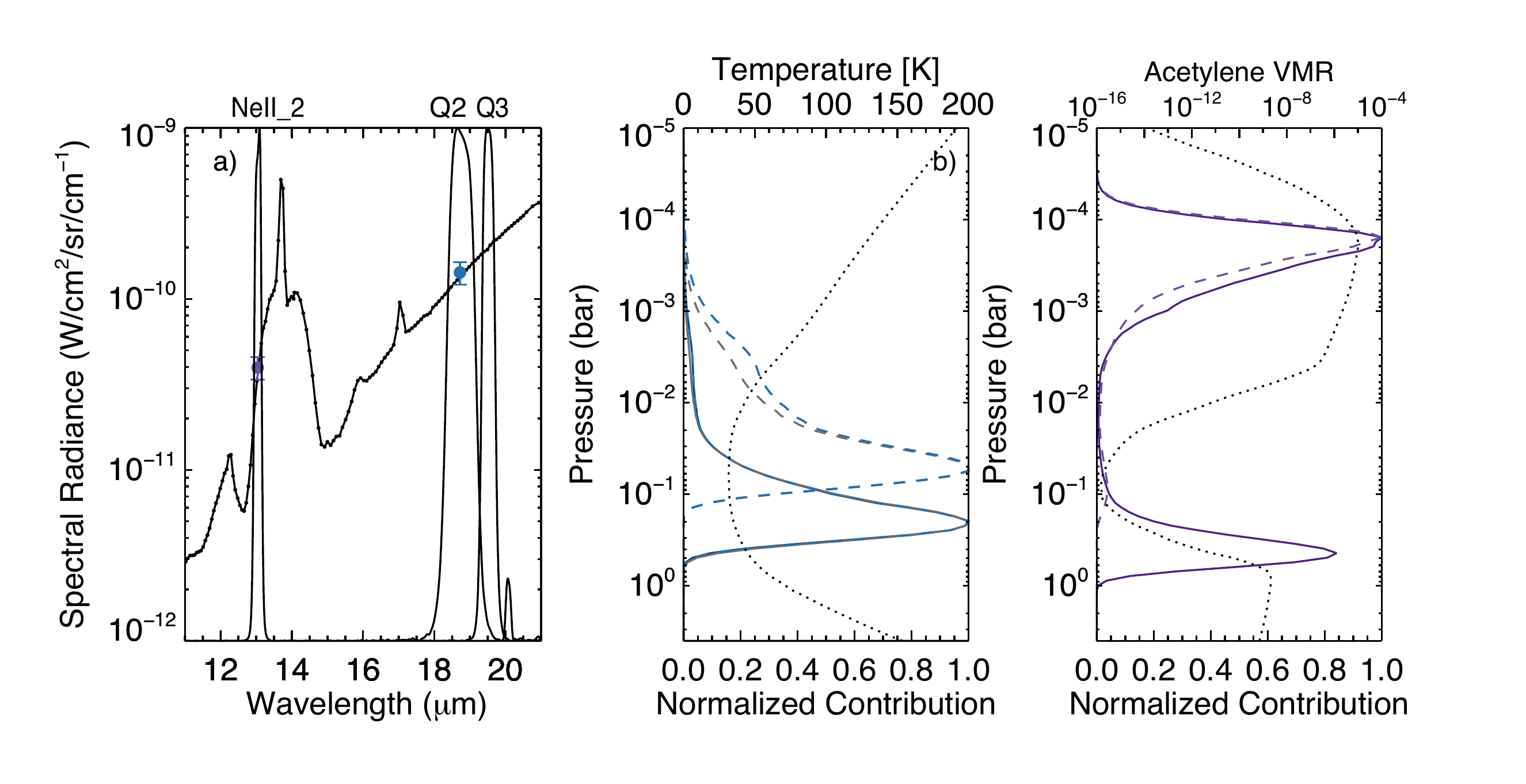}
\caption{a) 13.04-$\mu$m (NeII$\_$2), 18.72-$\mu$m (Q2), and 19.50-$\mu$m (Q3) normalized filter transmissions superimposed over the Short-Low Spitzer spectra of \cite{orton2014a}. The NeII$\_$2 filter senses emission from acetylene while the broader Q2 and Q3 continuum filters measure atmospheric temperatures via collision-induced hydrogen emission. Disk integrated values of the 2018 measurements are plotted with error bars representing a 20\% uncertainty. b) The normalized contributions for each filter superimposed at nadir (solid line) and limb (dashed). The middle panel shows the 18.72 $\mu$m (blue) and 19.50 $\mu$m (gray) contribution functions along with the temperature profile (dotted) of \cite{orton2014a} is shown for comparison. These Q-band filters probe the upper troposphere and tropopause, with the lowest pressures towards the limb. The right panel shows that contribution to the 13.04 $\mu$m (NeII$\_$2) emission peaks in stratosphere at all emission angles, although there is a significant contribution from the upper tropopause near nadir. The dotted line in this panel plots the globally averaged acetylene volume mixing ratio at equinox from \cite{orton2014a} and \cite{moses2018seasonal}. }
\label{fig:filters}
\end{figure}

Normalized transmissions and atmospheric contributions for the NeII$\_$2 and Q2 filters are shown in Figure \ref{fig:filters}. Filter transmissions were taken from the European Southern Observatory's (ESO) VISIR instrument webpage \footnote{https://www.eso.org/sci/facilities/paranal/instruments/visir/inst.html}. The emission observed through these filters is attributed to a range of atmospheric pressures dependent upon the atmospheric temperature, density, and opacity. These atmospheric contribution functions were evaluated for each filter as the functional derivatives of radiance with respect to temperature using the NEMESIS radiative-transfer suite \citep{irwin2008nemesis} and the atmospheric model of \cite{orton2014a}. 

\section{Correcting for Observational Effects}\label{appendix:b}
The observed radiance for each point on the planet's disk generally depends on the local emission angle ($\mu$) as determined by the time-dependent viewing geometry as seen from Earth.  Consequently,  differences in the observed emission angle alone lead to apparent changes in the observed radiance and a simple, direct comparison of the new and archival images was hindered by changes in the observing geometry over the intervening nine years.  
In addition, imperfect seeing and optical diffraction significantly reduced the observed radiance nearer to the limb of Uranus' relatively small disk, altering the observed center-to-limb variation in emission.  If this effect was not correctly accounted for, the retrievals from images would have yielded erroneously colder temperatures near the limb in our model of the true atmosphere (prior to diffraction and atmospheric distortion).  In theory, this blurring could have been corrected for by performing a mathematical deconvolution between the image and the effective point spread function (PSF); however, in practice, the process amplified the considerable noise in our images, overwhelming the signal in the absence of excessive smoothing, rendering the deconvolution profitless. 

In order to develop an accurate model of latitudinal temperature structure prior to blurring, we attempted to account for losses near the limb by first evaluating the effect of observed PSF on synthetic images and then using these results to create simple correction factors.  This was done as follows: Firstly, we extracted zonal averages of the radiances from the filtered images, avoiding points near the edge of the disk ($\mu$ $<$ 0.3). These zonally averaged radiances were then inverted to retrieve atmospheric temperatures or acetylene using an optimal estimation retrieval algorithm (NEMESIS \citep{irwin2008nemesis}). Beginning with initial profiles of atmospheric temperatures and acetylene based on \cite{orton2014a}, the inversion yielded optimized continuous profiles of atmospheric parameters as a function of pressure and latitude. For the latitudes corresponding to omitted points near the edge of the disk (i.e., points with emission angles greater than 72.5$^{\circ}$), values at the nearest sampled latitudes were used.  We then mapped these parameter profiles onto a disk (assuming zonal homogeneity), and computed the emerging radiances using NEMESIS, resulting in a synthetic image of the planet for each filter.  These synthetic images were convolved with appropriate PSFs (carefully determined from the corresponding stellar images) to yield synthetic blurred images.  By dividing the original synthetic images by the blurred synthetic images, we obtained simple, 2-dimensional factors that approximated how the convolutions altered the images.  We then applied these correction factors to the real data to approximately reconstruct the true center to limb variation in each image, before once again extracting zonal averaged radiances and retrieving an improved temperature structure.  Though this approach was not a technical deconvolution, we found that it adequately mimicked the general effect and ultimately allowed us to generate more accurate models of the observations. Though as noted, the limb-brightening near the very edge of the disk was too great in our modeling, indicating a need for improvement in future modeling. 

Modeling the images allowed us to characterize the different components that ultimately led to the final image (i.e., the signal, noise, diffraction, atmospheric blurring, viewing geometry) while yielding temperature and acetylene models that could be analyzed and remapped to different times and geometries. This approach also had the benefit of allowing us to compare and adapt models to the unaltered observations without the risk of corrupting the true data themselves by correcting for observational effects. The drawbacks are that the models are sensitive to assumptions regarding vertical profiles that cannot be constrained by our data. The anomalous limb brightening discussed in the text is likely a consequence of our assumed profile; however, in its failure, this modeling provides motivation for correcting the temperature profile in future work.

\section{Hints of Zonal Variability} \label{appendix:c}

\begin{figure}[ht!]
\includegraphics[width=\linewidth, trim=0in 0in 0in 0in, clip]{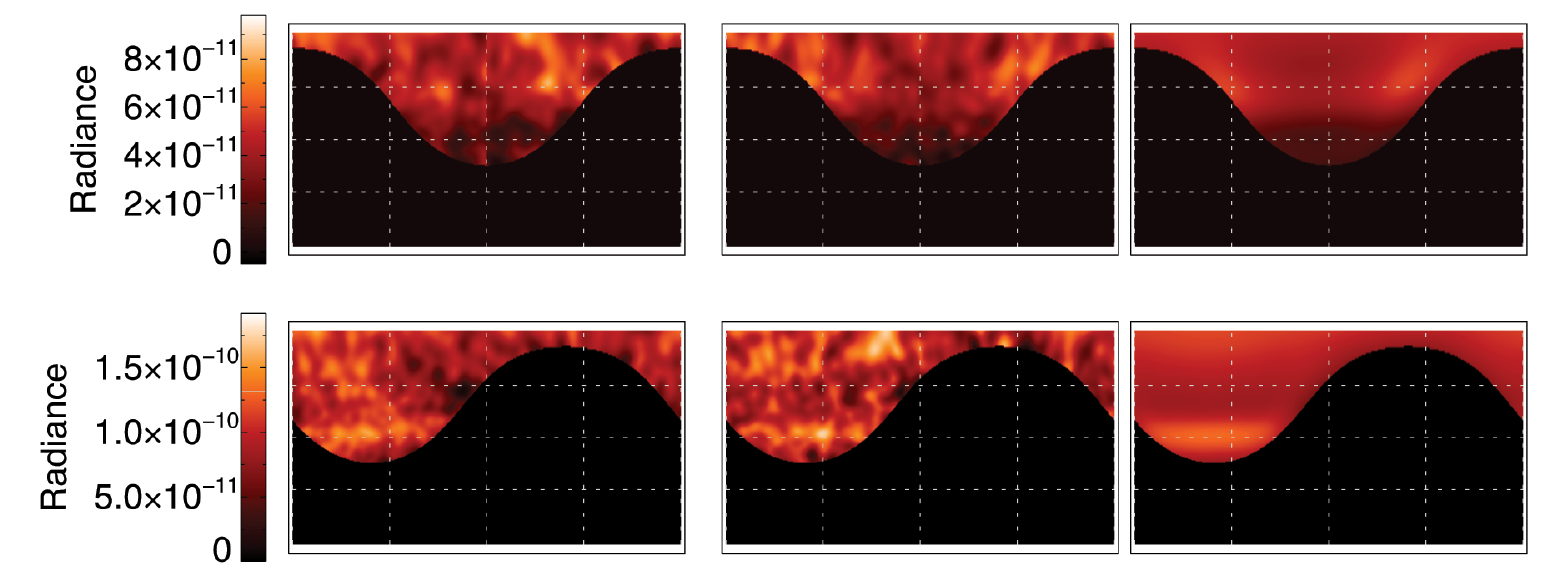}
\caption{Comparison of data and models, illustrating the contribution of random noise, which can essentially account for all the zonal variation across the disk. The top row represents 13 $\mu$m and the bottom row is 18.7 $\mu$m. The left column depicts maps of real data; the middle column shows a blurred synthetic images with random noise added to simulate the effect of statistical noise in the data; and the right column shows maps of the synthetic blurred image prior to the addition of synthetic noise. Radiances in the color bar are in units of $W/cm^2/sr/cm^{-1}$.
\label{fig:2018maps}}
\end{figure}

\begin{figure}[ht!]
\includegraphics[width=\linewidth, trim=.45in 2in .45in 2in, clip]{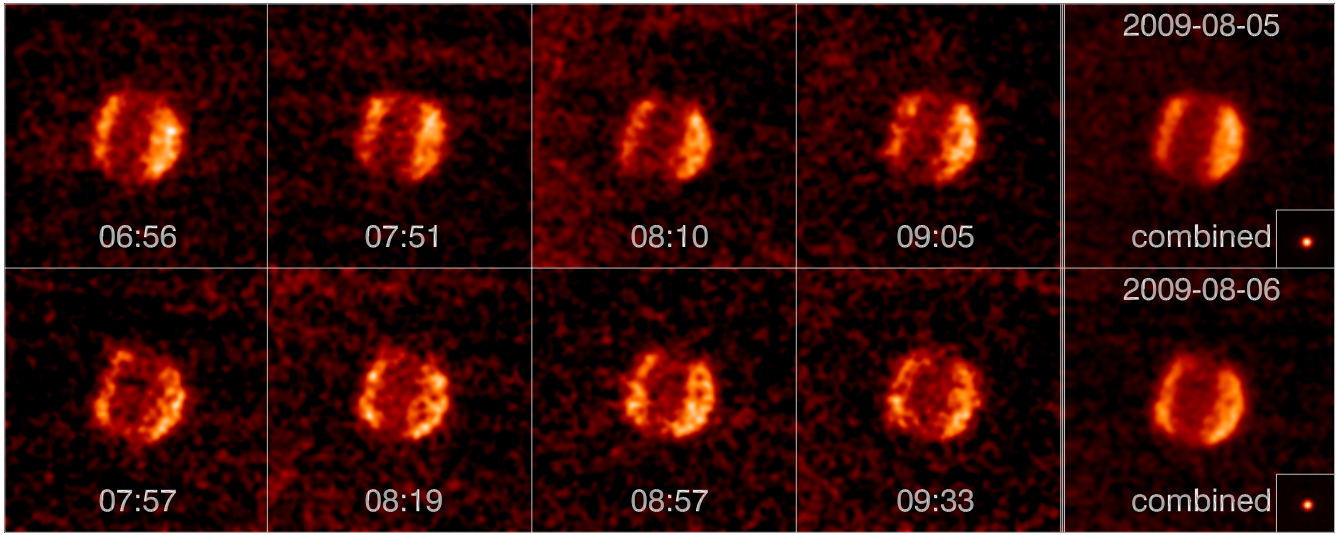}
\caption{August 5, 2009 (top) and August 6, 2009 (bottom) images in the NeII\_2 [13.04 $\mu$m] filter from from VLT-VISIR. Each row represents a multiple observations over a single night, corresponding to the times indicated and listed in Table \ref{table:equitable}, with the nightly means (smearing roughly 2 hrs of rotation) in the rightmost column.  Images shown here have been low-pass filtered with a 1-pixel standard deviation Gaussian blur for image clarity. Insets show the stars representative of the average spatial resolution of the seeing disc. The brightest regions in the two nightly means occur at adjacent longitudes, possibly suggesting zonal variation beyond the noise. For our analysis, we chose to average both nights together (see  Fig \ref{fig:equiobs}).}
\label{fig:neii2009}
\end{figure}

All of the zonal variation evident in the individual 2018 images are consistent with image noise, as seen by comparing the actual data to the modelled maps using realistic noise (computed from the standard deviation of the background skies). This is true for the averaged images as well as the individual images.  For example, Figure \ref{fig:2018maps} shows the mapped individual images along side synthetic maps created from our model. This justifies the choice to average individual images together given the low SNR. However, it is worth noting the brightest features in the NeII$\_$2 intriguingly appear at the similar latitudes ($\sim$40-45$^{\circ}$) and could constitute real features at the $\sim$2.5-$\sigma$ level. The 2009 13-$\mu$m images also reveal hints of possible zonal variability with coherently greater emission towards similar longitudes (see Figure \ref{fig:neii2009}, but these are marginally consistent with the level of distortion due to image noise and blurring. Zonal variability in the 13-$\mu$m emission would help explain variations between different longitudes noted in the disk-averaged \textit{Spitzer} observations \citep{orton2014a}. Investigating possible discrete physical features and zonal variability in the stratosphere will require greater SNR imaging in the future. 

\section{Variations in Time Due to Changes in Observing Geometry}
Finally, regardless of the cause of the seasonal cloud variability, decades of disk-integrated visible albedos (472 nm and 551 nm) show an asymmetric seasonal pattern that cannot be explained by viewing geometry alone \citep{lockwood2019final}. Although these historical measurements do not yet cover a complete cycle to evaluate changes in solstices, they do show a clear asymmetry between equinoxes.  The 1946 southern vernal equinox has a significantly lower albedo than the 2007 northern vernal equinox, reaching seasonal minimum values shortly after each equinox. We can attempt to compare this pattern of observed visible albedo to expected variation in the thermal emission.

\begin{figure}[ht!]
\includegraphics[width=\linewidth, trim=.0in 0in .0in 0in, clip]{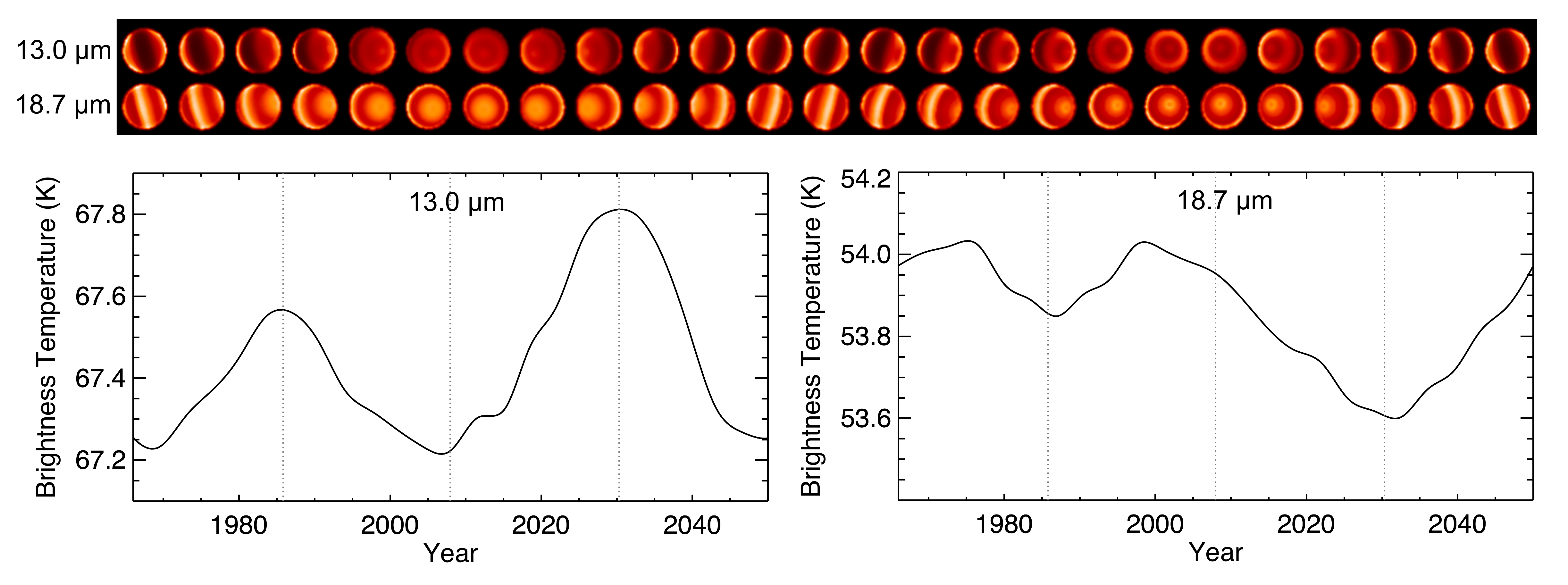}
\caption{Variations in the radiance due to differences in observing geometry between the 1966  and 2050 equinoxes at intervals of 3.5 yrs. (top) Synthetic images of thermal emission at 13 $\mu$m (top row) and 18.7 $\mu$m (bottom row) produced by forward modeling our hybrid model of temperatures from 2018, 2009, and Voyager. Images show the view from Earth at the 3.5 year intervals with the disk size normalized to correct for secular variations from Earth's orbit; these secular variations were accounted for in our disk-integrated brightness. (bottom) Disk integrated brightness temperatures for 13 $\mu$m (left) and 18.7 $\mu$m (right) produced from a spline fit to the integrated synthetic images. The assumed asymmetry is evident, but variations are still small relative to typical calibration uncertainties.}
\label{fig:oneyear}
\end{figure}

Combining retrieved temperatures from Voyager \citep{orton2015thermal} with 2009 and 2018 temperatures retrieved from imaging, we developed a global model of temperatures. This was then forward-modeled over a range of viewing geometries encompassing a full orbit of Uranus as seen from Earth (see Fig \ref{fig:oneyear}). The model assumed the asymmetry in 2009 imaging was invariant. The variations we see are thus asymmetric in time, although the south polar regions are purely speculative and assumed to be symmetric with the north at 13 $\mu$m and consistent with Voyager at 18 $\mu$m. However, the vernal and autumnal equinoxes are symmetric, unlike what is seen in the visible \citep{lockwood2019final}, so these thermal curves provide no obvious clues to the unexplained visible asymmetry. 

Given discrepancies between the data and photochemical model, we neglected seasonal variations in the photochemical abundances \citep{moses2018seasonal}, opting instead to use the retrieved abundances held fixed over the year. 

From a practical observational perspective, the range in brightness temperatures are subtle and probably undetectable in the published record of observations given larger uncertainties in calibration at these wavelengths, which may be as large as 30\% in radiance.  For comparison, \cite{orton1987spectra} measured Uranus brightness temperatures in 1985 of $\sim$63 K at 13 $\mu$m and just greater than 53 K at 18.7 $\mu$m, both of which fall below the plotted ranges of our annual projections.  

\end{document}